\documentclass[final]{IEEEtran}
\usepackage{cite}
\usepackage{amsmath,amssymb,amsfonts,amsthm}
\usepackage[utf8]{inputenc}
\usepackage[T1]{fontenc}
\usepackage{algorithm}
\usepackage{algorithmic}
\usepackage{graphicx}
\usepackage{textcomp}
\usepackage{xcolor}
\usepackage{url}
\usepackage{lipsum}
\usepackage{multirow}
\usepackage[normalem]{ulem}
\usepackage{bbding}
\usepackage{pifont}
\usepackage{wasysym}
\usepackage{bbm}
\usepackage{dsfont}
\usepackage{mathtools, nccmath}
\usepackage{makecell}
\usepackage{aligned-overset}
\usepackage{stfloats}
\usepackage{float}
\usepackage{soul}
\usepackage{threeparttable}
\usepackage{xfrac}
\ifCLASSOPTIONcompsoc
\usepackage[caption=false, font=normalsize, labelfont=sf, textfont=sf]{subfig}
\else
\usepackage[caption=false, font=footnotesize]{subfig}
\fi
\newfloat{procedure}{tbp}{lop}
\floatname{procedure}{Procedure}

\DeclareMathAlphabet{\mathbcal}{OMS}{cmsy}{b}{n}
\def\BibTeX{{\rm B\kern-.05em{\sc i\kern-.025em b}\kern-.08em
		T\kern-.1667em\lower.7ex\hbox{E}\kern-.125emX}}
\newtheorem{remark}{Remark}

\begin{document}
	\title{SARA: SLO-Aware Resource Allocation for Disaggregated Agentic LLM Services%
	}
	\author{Shicong~Liu,~\IEEEmembership{Graduate~Student~Member,~IEEE}, Xianghao~Yu,~\IEEEmembership{Senior~Member,~IEEE}, 
	Zhen~Gao,~\IEEEmembership{Member,~IEEE},
	and Jun~Zhang,~\IEEEmembership{Fellow,~IEEE}
		\thanks{
			({\it Corresponding Author: Xianghao Yu.})
			
			Shicong Liu and Xianghao Yu are with the Department of Electrical Engineering, City University of Hong Kong, Hong Kong (email: sc.liu@my.cityu.edu.hk, alex.yu@cityu.edu.hk).
			
			Zhen Gao is with the Advanced Research Institute of Multidisciplinary Science (ARIMS), Beijing Institute of Technology. 
			
			Jun Zhang is with the Department of Electronic and Computer Engineering, The Hong Kong University of Science and Technology, Hong
			Kong (email: eejzhang@ust.hk)
		}
	}
	\maketitle
	\begin{abstract}
		Recent advances in large language models (LLMs) are driving the emergence of multi-modal and agentic services {for mobile users through cloud and edge infrastructures}, where long-context workloads pose daunting challenges 
		for inference latency. 
		Existing disaggregated LLM serving systems largely rely on hardware profiling, configuration enumeration, or heuristic scheduling, offering limited analytical guidance for cost-efficient resource allocation. 
		In this paper, we propose SARA, a \uline{S}ervice level objectives (SLOs)-\uline{A}ware \uline{R}esource \uline{A}llocation framework for disaggregated agentic LLM serving systems, which maximizes goodput under a deployment cost constraint and a series of quantile-based SLO constraints. By capitalizing on queuing theory, we first model the prefill, KV cache transfer, and decode stages as an M/G/$k_{\rm p}$ queue, an M/G/$1$ queue, and a generalized birth-death process, respectively. 
		The analysis 
		reveals that the prefill and decode stages are dominantly limited by computational capacity and high-bandwidth memory (HBM) resources, respectively. 
		With these mathematical models, we further derive tractable tail behaviors of different stage-wise service level metrics for both light- and heavy-tailed workloads. 
		These characterizations explicitly map workload, model architecture, and hardware parameters to stage-wise SLO constraints and minimum resource requirements.
		{We also reveal that the required numbers of prefill and decode devices scale quadratically with the input and output lengths, respectively, and that stringent SLOs and heavy-tailed workloads substantially amplify the resource requirements beyond the stability condition.}
		Finally, we develop an effective resource allocation framework to maximize system goodput under limited cost budgets.
		Simulation and hardware results demonstrate that the proposed framework accurately predicts the stage-wise SLO with mean errors below $\bf 5\%$, and improves system goodput by $\bf 26.6\%$ on average over state-of-the-art baseline methods under the same deployment cost.
	\end{abstract}
	\section{Introduction}
	\bstctlcite{IEEEexample:BSTcontrol}\label{sec:intro}
	Large language models (LLMs) are rapidly evolving from general-purpose conversational tools into foundational services for intelligent web search, code generation, multi-modal understanding, complex reasoning, and autonomous agentic workflows~\cite{11540994,11175596,11474254}. As model capabilities and context windows continue to expand, LLM applications are moving beyond single-turn dialogue interactions toward multiple-turn agentic tasks involving high-effort reasoning~\cite{3600270.3602070} and decision making~\cite{10970024}. 
	Unlike model training, online LLM inference must continuously serve large and fluctuating workloads under stringent service level objectives (SLOs), {including quantile-based latency metrics such as time to first token (TTFT) and time per output token (TPOT).} 
	Consequently, inference is becoming a persistent and growing infrastructure workload, and its resource efficiency will largely determine whether LLM services can be deployed at scale with sustainable cost~\cite{IEA2026EnergyAI}.
	
	The resource demands of LLM inference vary substantially across its execution process~\cite{distserve,ModServe,prefill,patel2024splitwise,10946802}. Upon receiving a user request, which typically contains textual or multi-modal input data, the model first tokenizes and embeds the entire input into large-scale matrices for subsequent intensive matrix multiplication. This stage is commonly referred to as the \textit{prefill} stage, which requires substantial computational resources. After the prefill stage, the model starts the autoregressive token generation process, i.e., the \textit{decode} stage. This stage contains less parallel computation but frequent access to the intermediate results, which are stored in the \textit{key and value (KV) cache} computed in the prefill stage. Therefore, the decode stage requires excessively high memory bandwidth.
	
	The heterogeneous resource requirements of the prefill and decode stages pose significant challenges for the design of LLM inference at data centers. 
	When the two stages are executed within the same resource pool, they will interplay with each other, significantly elongate inference time~\cite{distserve}, and reduce inference efficiency. 
	Besides, colocated deployment reduces resource utilization and increases the cost of data center expansion as the number of user requests increases. 
	To mitigate interference between the two stages and better match resources to their respective demands, a disaggregated prefill and decode (PD) deployment strategy has become an important design direction for large-scale LLM serving systems~\cite{distserve,prefill,patel2024splitwise,ModServe}. {Moreover, such disaggregation paves the way for edge-assisted LLM services for mobile users. The compute-intensive prefill is isolated in data centers, while the memory-intensive decode stage can be pushed toward edge nodes in proximity to users.}
	
	However, the disaggregated LLM architecture also introduces several novel problems. 
	First, the decode stage relies on the KV cache generated during the prefill stage. When the two stages are deployed separately, the KV cache must be transferred from the prefill to decode resource pool, making the transfer latency an important factor in inference efficiency. Second, it further introduces the problem of how resources should be provisioned and allocated across the disaggregated resource pools and communication link of the KV cache transfer. Finally, the heavy-tailed user request distributions typically encountered in agentic and multi-modal workloads further complicate performance analysis and resource planning. 
	These challenges call for an analytical framework and resource allocation scheme for disaggregated LLM serving systems.

	\subsection{Related Work}
	
	To meet increasingly stringent SLO requirements in LLM serving systems, research efforts have been devoted to improving LLM inference efficiency. 
	From the perspective of model architecture, multi-query attention (MQA) and grouped-query attention (GQA) were proposed to reduce the computational complexity and memory access overhead by allowing multiple query matrices to share KV matrices~\cite{MQA,GQA}. Besides, 
	mixture-of-experts (MoE) models integrate multiple smaller domain-specific models into a unified architecture, employing dynamic gating based on the input user requests~\cite{moe}. Since only a sparse subset of these experts is activated during inference, the MoE architecture significantly enhances the inference efficiency. 
	Furthermore, to accommodate the heterogeneous resource demands of Transformer-based LLMs, Sarathi-Serve partitions long user requests into smaller chunks and flexibly schedules their execution to improve resource utilization~\cite{3691938.3691945}. 
	Beyond fine-grained execution scheduling, several recent works have improved inference efficiency through workload-aware resource configuration and request routing. 
	Specifically, DynamoLLM configures resource pools with different hardware compositions and routes requests according to their sequence characteristics to improve energy efficiency~\cite{10946802}. ModServe dispatches modality-specific workloads to specialized resource pools to enhance resource utilization and inference throughput~\cite{ModServe}. In addition, Taiji dynamically adapts resource allocation and request scheduling according to the current SLO bottleneck to increase SLO achieving throughput~\cite{wang2026taiji}. 

	However, these approaches still colocate the prefill and decode stages on shared resource pools and thus cannot fundamentally eliminate their mutual interference. 
	Moreover, even when scheduling mitigates such interference to some extent, the shared deployment couples the prefill and decode resources, limiting the independent scaling of the two stages according to their distinct workload characteristics and SLO bottlenecks.

	To eliminate the mutual interference between prefill and decode stages and further facilitate resource planning, recent studies have focused on the disaggregated LLM inference architecture, which executes the two stages on separate resource pools~\cite{patel2024splitwise,distserve,10.1145/3773772,prefill}. In particular, Splitwise is one of the pioneer works that implements an LLM serving system onto separate prefill and decode resource pools, while still preserving a mixed resource pool for elastic workload balance~\cite{patel2024splitwise}. Moreover, 
	DistServe then proposed to maximize the goodput, i.e., the throughput achieving designated SLO requirements, of the inference system, where the parallelism strategy and resource allocation are achieved via enumeration over a reduced search space~\cite{distserve}. 
	Furthermore, Mooncake develops a KV cache-centric disaggregated architecture and improves serving performance through carefully designed cache scheduling and transfer~\cite{10.1145/3773772}. 
	In addition, PaaS leverages linear-complexity attention to reduce KV cache transfer overhead and introduces scheduling policies that jointly exploit cache availability and communication bandwidth~\cite{prefill}.  

	Despite these advances, existing disaggregated systems still rely primarily on hardware performance profiling~\cite{patel2024splitwise}, configuration enumeration~\cite{distserve}, or heuristic scheduling~\cite{10.1145/3773772,prefill}. Although effective for particular platforms and workloads, these approaches offer limited theoretical guidance for offline data-center capacity planning or low-overhead online reconfiguration. 
	In particular, they lack a tractable model that maps workload statistics and heterogeneous resource allocations to stage-wise SLO-attainment probabilities and minimum required resources. 
	These gaps motivate our SLO-aware system modeling and characterization, as well as cost-constrained resource allocation framework for disaggregated LLM serving.
	
	\subsection{Contributions}
	
	In this paper, we investigate SLO-aware resource allocation for disaggregated LLM serving systems to maximize the system goodput under cost constraints. 
	The main contributions are summarized as follows.
	
	\begin{itemize}
		\item To capture the heterogeneous resource demands of disaggregated LLM serving systems, we develop stochastic mathematical models including an M/G/$k$ prefill queue, an M/G/$1$ KV cache transfer queue, and a continuous batching decode process model. 
		The models explicitly incorporate key system parameters, and provide mathematical insights on the resource demands. 
		Based on these models, we further derive the transition conditions between the compute- and memory-intensive regimes, validating that prefill and decode stages are predominantly constrained by computational capacity and memory bandwidth, respectively. 
		\item We derive tractable and analytical characterizations of the stage-wise latency quantiles, including TTFT, KV cache transfer latency, and TPOT to move beyond conventional empirical SLO assessments. 
		Specifically, under limited cost constraint, heavy-traffic approximation is employed to characterize the waiting and sojourn latency for both the prefill and KV cache transfer queuing models, while the moment matching method is leveraged to characterize the latency quantile in the decode stage. 
		The resulting analytical expressions explicitly connect system parameters and hardware resources to SLOs under both light- and heavy-tailed workloads.
		
		\item We further propose a resource allocation framework to solve the goodput maximization problem for disaggregated LLM serving systems. In particular, stage-wise minimum resource requirements can be obtained directly from the derived closed-form expressions connecting SLOs and hardware demands, and the maximum goodput is obtained via a bisection search. 
		\item We finally conduct practical hardware measurements and simulations to validate the proposed analytical framework. The results verify the accuracy of the derived stage-wise latency quantiles with representative model architectures and both light- and heavy-tailed workloads, demonstrating the applicability of the proposed SLO-aware resource allocation framework.
	\end{itemize}
	
	\par {\it Notations}: We use lowercase and uppercase boldface letters to denote vectors and matrices, respectively. 
	The element at the $k$-th row and the $m$-th column of matrix $\bf H$ is denoted as ${\bf H}[k,m]$, and the $n$-th element in the vector $\bf h$ is denoted by ${\bf h}[n]$. $\{{\bf H}_n\}_{n=1}^N$ represents a matrix set with the cardinality of $N$. The superscripts $(\cdot)^{T}$ and $(\cdot)^{H}$ represent the transpose and conjugate transpose operators, respectively
	, and $\det(\cdot)$ denotes the determinant operator. 
	$\mathbb{R}$ and $\mathbb{Z}$ denote the sets of real numbers and integers, respectively. We use ${\mathcal{B}}[n,p]$ and ${\mathcal{LN}}[\mu, \sigma^2]$ to denote binomial distribution with parameters $\{ n,p \}$ and log-normal distribution with parameters $\{ \mu,\sigma^2 \}$, respectively. We use $\mathbb{E}[X]$ and ${\rm Var}(X)$ to denote the mean and variance of random variable $X$.

	\section{Preliminaries}\label{sec:pre}
	In this section, we introduce basic concepts of a typical LLM inference system.

	\subsection{Generative Pre-Trained Transformer}\label{sec:gpt}
	
\begin{figure*}[t]
		\centering
		\includegraphics[width=\textwidth]{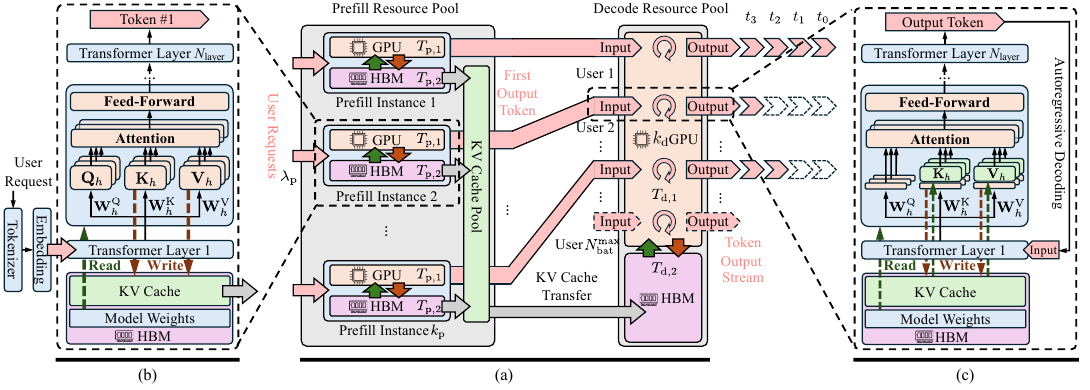}
		\caption{(a) Overall disaggregated LLM inference architecture. Key computational operations of one user request in the (b) prefill stage and (c) decode stage~\cite{shi2026multisegmentattentionenablingefficient}.}
		\label{fig:prel}
	\end{figure*}

	The generative pre-trained Transformer (GPT) has emerged as the dominant architecture for modern LLMs~\cite{gpt,10500411}. It is first pre-trained on massive text-based or multi-modal corpora~\cite{NEURIPS2023_6dcf277e,gemini15} to predict the next token given preceding tokens, and can then be adapted or aligned for various downstream applications. Most LLMs adopt a decoder-only, multi-layer Transformer architecture~\cite{grattafiori2024llama3herdmodels,qwen3,NEURIPS2020_1457c0d6}, where each layer comprises a self-attention module and a feed-forward network~\cite{NIPS2017_3f5ee243}, as illustrated in Fig.~\ref{fig:prel}. During a typical LLM inference process, the core computational operations are as follows:
	\begin{itemize}
		\item {\bf Tokenization \& Embedding}: A user request typically involves textual or other multi-modal input data. They are first {tokenized into a sequence with $L_{\rm i}$ input tokens}, which are then mapped to matrices $\mathbf{X} \in \mathbb{R}^{L_{\rm i} \times N_{\mathrm{mod}}}$ via an embedding lookup table.
		\item {\bf Attention}: The input matrix ${\bf X}$ experiences large-scale multiplications. Take classical multi-head attention (MHA)\footnote{In modern LLMs, the attention module consists of hybrid structures, i.e., full attention~\cite{NIPS2017_3f5ee243,GQA,MQA} and linear-complexity attention modules~\cite{kimilinear,SWA,openai2025gptoss120bgptoss20bmodel}, which reduce the computational complexity and memory consumption.}~\cite{NIPS2017_3f5ee243} as an example, the input matrix is first mapped to $N_{\rm head}$ query and KV matrices in shape $L_{\rm i}\times N_{\rm attn}$ by 
		\begin{equation}
			{\bf Q}_h = {\bf X}{\bf W}^{Q}_h,~~
			{\bf K}_h = {\bf X}{\bf W}^{K}_h,~~
			{\bf V}_h = {\bf X}{\bf W}^{V}_h,
			\label{eq:qkv}
		\end{equation}
		respectively, 
		where $1\leq h \leq N_{\rm head}$ denotes the index of attention heads, and ${\bf W}^{Q}_h$, ${\bf W}^{K}_h$, and ${\bf W}^{V}_h$ in shape $N_{\rm mod} \times N_{\rm attn}$ are the pre-trained weight matrices. Each head then performs scaled dot-product attention in parallel as%
		\begin{equation}
			\!{\bf A}_{\rm H}^{(h)}={\rm softmax}\left( \frac{{\bf Q}_h {\bf K}_h^T}{\sqrt{N_{\rm attn}}} \!+\!{\bf M}\right) {\bf V}_h\in\mathbb{R}^{L_{\rm i} \times N_{\rm attn}},\label{eq:attn}
		\end{equation}
		where ${\bf M}$ in shape $L_{\rm i}\times L_{\rm i}$ denotes the causal mask matrix, whose upper-triangular entries are set to $-\infty$, while the diagonal and lower-triangular entries are set to zero. This mask prevents each token from attending to future tokens during self-attention, and reduces the computational complexity by two-fold in ideal cases.
		The outputs of all heads are concatenated and multiplied by a pre-trained output projection matrix ${\bf W}_{\rm O}$ as
		\begin{equation}
			{\bf A}_{\rm H} = \left[ {\bf A}_{\rm H}^{(1)}, \cdots, {\bf A}_{\rm H}^{(N_{\rm head})} \right] {\bf W}_{\rm O}\in\mathbb{R}^{L_{\rm i} \times N_{\rm mod}}.
			\label{eq:agg}
		\end{equation}
		\item {\bf Feed-Forward}: The matrix ${\bf A}_{\rm H}$ is further processed by large-scale matrix operations involving two matrix multiplications and a non-linear activation function as
		\begin{equation}
			{\bf O} = \sigma\left( {\bf A}_{\rm H} {\bf W}_1 \right) {\bf W}_2 \in\mathbb{R}^{L_{\rm i} \times N_{\rm mod}},
			\label{eq:ff}
		\end{equation}
		where ${\bf W}_1\in\mathbb{R}^{N_{\rm mod}\times N_{\rm hid}}$ and ${\bf W}_2\in\mathbb{R}^{N_{\rm hid}\times N_{\rm mod}}$ are the feed-forward projection matrices. Output matrix ${\bf O}$ acts as the input matrix ${\bf X}$ of the next Transformer layer. 
		{After the last Transformer layer, the last row vector of ${\bf O}$ is projected and normalized into a probability distribution, where the first output token is generated via sampling or greedy search~\cite{pmlr-v202-leviathan23a}.}
		\item {\bf Normalization}: Normalization is applied to each row of ${\bf O}$ (post-norm) or ${\bf X}$ (pre-norm) independently to ensure numerical stability among layers.
	\end{itemize}  
	
	The computational complexity of Transformers is mainly determined by general matrix multiplications (GEMMs)~\cite{distserve}. 
	The number of multiplications in one Transformer layer is characterized as follows. The projection operations in~\eqref{eq:qkv}, attention operation in~\eqref{eq:attn}, the aggregation operation in~\eqref{eq:agg}, and the feed-forward in~\eqref{eq:ff} require $3L_{\rm i}N_{\rm mod}^2$, $\left( L_{\rm i}^2\!+\!L_{\rm i} \right) N_{\rm mod}$, $L_{\rm i}N_{\rm mod}^2$, and $2L_{\rm i}N_{\rm mod}N_{\rm hid}$ multiplications in the ideal case, respectively. Therefore, the overall number of multiplications for $N_{\rm layer}$ layers is given by
	\begin{equation}
		\begin{aligned}
			&N_{\rm mul}(L_{\rm i})\\
			={}&N_{\rm layer}\!\left( 4L_{\rm i}N_{\rm mod}^2 \!+\! ( L_{\rm i}^2\!+\!L_{\rm i} ) N_{\rm mod} \!+\! 2L_{\rm i}N_{\rm mod}N_{\rm hid} \right).
		\end{aligned}
		\label{eq:nmul}
	\end{equation}
	Note that for advanced attention variants such as multi-query attention (MQA)~\cite{MQA} and grouped-query attention (GQA)~\cite{GQA}, $N_{\rm KV}\leq N_{\rm head}$ key-value heads are shared among $N_{\rm head}$ query heads in the attention operation~\eqref{eq:attn}. Let $g = N_{\rm KV}/N_{\rm head}\in (0,1]$ be the compressive factor and then the only difference in $N_{\rm mul}$ occurs in the projection operations of~\eqref{eq:qkv} as $(1+2g)L_{\rm i}N_{\rm mod}^2$.
	
	\subsection{Disaggregated LLM Inference Architecture}\label{sec:dis}
		
	As introduced in Sec.~\ref{sec:intro}, resource requirements vary significantly across the execution stages of LLM inference. Specifically, generating the first token requires a full execution of all Transformer layers, resulting in excessively high computational demand. During the subsequent autoregressive token generation, the intermediate KV matrices can be reused, which substantially reduces computational operations. However, frequent access to the KV cache makes memory bandwidth the major demand. 
	Consequently, an inference process can be divided into three disaggregated and sequential stages according to their resource demands as follows.
	
	\begin{itemize}
		\item {\bf Prefill}: The prefill stage spans from user prompt input to the generation of the first output token. It includes intensive attention and feed-forward operations across all transformer layers, as shown in Fig.~\ref{fig:prel}(b) and introduced in Sec.~\ref{sec:gpt}.
		Note that the KV matrices  $\{ {\bf K}_h \}_{h=1}^{N_{\rm KV}}$ and $\{ {\bf V}_h \}_{h=1}^{N_{\rm KV}}$ associated with attention operations are stored as the KV cache to support subsequent stages, while $\{ {\bf Q}_h \}_{h=1}^{N_{\rm head}}$ are discarded after the first token is generated. 
		Owing to the intensive large-scale matrix multiplications involved, the resource bottleneck of this stage is the computational capacity~\cite{distserve}.
	
		\item {\bf KV Cache Transfer}: The KV matrices $\{ {\bf K}_h \}_{h=1}^{N_{\rm KV}}$ and $\{ {\bf V}_h \}_{h=1}^{N_{\rm KV}}$ computed during the prefill stage remain fixed and are reused throughout the inference. Therefore, in the disaggregated LLM inference framework, these matrices are stored as KV cache and must be further transferred to the decode resource pool to accelerate further computations. 
		\item {\bf Decode}: In this stage, the LLM autoregressively generates a token stream based on the entire KV cache\footnote{The decode stage also produces incremental KV cache when generating new output tokens, which is appended to the total KV cache for subsequent inference, as illustrated in Fig.~\ref{fig:prel}(c). Yet, the additional computational overhead introduced here is usually negligible compared with the entire prefill stage.}. Specifically, as illustrated in Fig.~\ref{fig:prel}(c), {in each iteration of the decode stage, the previous token acts as the input and} 
		yields incremental query and KV vectors, which together with the KV cache finish the following attention operations to generate the next token. 
		The decode iteration ends when an end-of-sequence (EOS) token is generated or the maximum output length is reached. 
		Because of frequent memory accesses of the KV cache, this stage requires extremely high memory bandwidth for efficient inference.
	\end{itemize}
	
	A typical disaggregated LLM inference process is illustrated in Fig.~\ref{fig:prel}. 
	Specifically, each prefill resource pool employs $k_{\rm p}$ prefill instances, which process incoming user requests in parallel. Upon completion, KV caches and the first output tokens of user requests are transferred to the decode resource pool, where output token streams are autoregressively generated in parallel via continuous batching~\cite{280922}.

	\subsection{KV Cache}\label{sec:kv}
	
	As we have introduced in Sec.~\ref{sec:dis}, KV cache is the intermediate output mainly generated in the prefill stage and reused during the following LLM inference procedure. In the decode stage, it stores KV matrices computed by each Transformer layer {using} all previously input and output tokens, so that the subsequent autoregressive iterations can directly reuse them without repeated computations. 
	
	However, for LLMs possessing several hundred billion of parameters, the resulting KV cache can also be extremely memory-consuming. For example, in an MHA-dominated architecture such as GPT-3 175B~\cite{NEURIPS2020_1457c0d6}, the typical KV cache memory consumption with $L_{\rm i} = 2048$ input tokens and $N_{\rm bits} = 16$ bits quantization in the prefill stage is given by
	\begin{equation}
		\begin{aligned}
			\Omega_{\rm KV}^{\rm MHA} (L_{\rm i}) = \frac{2 L_{\rm i} N_{\rm layer} N_{\rm mod} N_{\rm bits} }{8\times 1024^3} = \gamma L_{\rm i}= 9~({\rm GiB}),
			\label{eq:sizekv}
		\end{aligned}
	\end{equation} 
	where GPT-3 possesses $N_{\rm layer}=96$ layers of Transformers with model dimension $N_{\rm mod} = 12,288$, and $\gamma = N_{\rm layer}N_{\rm mod}N_{\rm bits}/4/1024^3$ is the KV cache scaling factor. Similarly, for GQA and MQA attention variants, the memory footprint is $\Omega_{\rm KV} (L_{\rm i}) = g\gamma L_{\rm i}$.
	
	With the emergence of hybrid-attention architecture, where the MHA is replaced by the mixture of a small number of full-attention layers and a larger number of linear-complexity or bounded-state layers~\cite{prefill,NIPS2017_3f5ee243,MQA,GQA,openai2025gptoss120bgptoss20bmodel,kimilinear,SWA}, the memory consumption of the KV cache is substantially reduced, making the KV cache transfer across data centers possible.

	\section{System Model}
	\label{sec:system_model}

	In this section, we introduce the mathematical model developed for the disaggregated LLM service architecture.

	\subsection{M/G/$k_{\rm p}$ Queuing Prefill Model}\label{sec:pref}
	
	Consider a prefill resource pool that is composed of $k_{\rm p}$ identical prefill instances with replications of the same LLMs. {Each instance is equipped with GPU devices and high bandwidth memory (HBM) as shown in Fig.~\ref{fig:prel}(a)}, and serves one user request at a time. User requests waiting for available instances form a shared queue. 
	
	Based on the behavior of prefill operations, we model the prefill stage as a queuing system with $k_{\rm p}$ parallel servers. 
	Since the LLM service requests typically come from a large number of independent users, the arrival process can be treated as a Poisson process with arrival rate $\lambda$~\cite{3691938.3691945,distserve}. 
	For a given input sequence with $L_{\rm i}$ tokens at a prefill instance, according to~\eqref{eq:nmul}, 
	the computational latency of the prefill stage is calculated by
	\begin{equation}
		T_{\rm p,1}(L_{\rm i}) = \frac{N_{\rm mul}\left( L_{\rm i} \right) }{F_{\rm p}}, 
		\label{eq:Tp1}
	\end{equation}
	where $F_{\rm p}$ denotes the compute capacity of each prefill instance in the resource pool. 
	
	Besides computational operations, the prefill stage also introduces HBM traffic latency, which mainly consists of model weight loading and KV cache writing. Specifically, the memory footprints of the model weight and KV cache are given by $\Omega_{\rm LLM}$ and $\Omega_{\rm KV}(L_{\rm i})$ in~\eqref{eq:sizekv}, 
	respectively, which result in the memory prefill latency 
	\begin{equation}
		T_{\rm p,2}\left(L_{\rm i} \right) = \frac{\Omega_{\rm LLM}+\Omega_{\rm KV}(L_{\rm i}) }{B_{\rm HBM}},\label{eq:Tp2}
	\end{equation}
	where $B_{\rm HBM}$ is the bandwidth of HBM. Hence, the overall service time $T_{\rm p}$ is then modeled as
	\begin{equation}
		T_{\rm p}\left(L_{\rm i} \right) = \max\left( T_{\rm p,1}\left(L_{\rm i} \right), T_{\rm p,2}\left(L_{\rm i} \right) \right).%
		\label{eq:prefillTp}
	\end{equation}
	In summary, we model the prefill stage as an M/G/$k_{\rm p}$ queue~\cite{Kingman2009}. Prefill utilization is given by $\rho_{\rm p}= \lambda \mathbb{E}[T_{\rm p}(L_{\rm i})]/k_{\rm p}$, which must satisfy $\rho_{\rm p}<1$ to ensure the stability of the queue.
	
	\subsection{KV Cache Transfer}
	
	After the prefill stage, the generated KV cache must be transferred to the decode resource pool. 
	{The departure traffic from the $k_{\rm p}$ prefill instances is forwarded to the KV cache transfer stage. When $k_{\rm p}$ is large and the departure traffics from $k_{\rm p}$ instances are mutually independent, the superposed departure process of the KV cache can also be approximated as a Poisson process~\cite{kingman1992poisson}. For a stable prefill queue without request loss or abandonment, the KV cache transfer arrival rate $\lambda_{\rm KV}$ equals the user request arrival rate of the prefill stage as $\lambda_{\rm KV} = \lambda$.}
	Let $B$ denote the link bandwidth for KV cache transfer, the service time is given by 
	\begin{equation}
		T_{\rm TR} \left(L_{\rm i}\right) = \frac{\Omega_{\rm KV}\left( L_{\rm i} \right)}{B} = \frac{g\gamma L_{\rm i}}{B}.\label{eq:TKV}
	\end{equation}
	Therefore, the KV transfer stage can also be modeled as an M/G/$1$ queue, and the utilization is 
	\begin{equation}
		\rho_{\rm KV} = \lambda \mathbb{E}[T_{\rm TR}(L_{\rm i})] = \frac{\lambda \mathbb{E}[\Omega_{\rm KV}(L_{\rm i})]}{B}<1.\label{eq:rhoKV}
	\end{equation}

	\subsection{Generalized Birth-Death Decode Model}
	\label{sec:dec}
	
	As shown in Fig.~\ref{fig:prel}(c), in the decode stage, the next output token is generated based on the input and all previous output tokens. 
	The reuse of KV cache significantly alleviates the computational overhead of token generation. To fully exploit the computational capability of the decode resource pool, requests containing the first output tokens generated by the prefill stage are executed via continuous batching~\cite{280922}. This enables the sharing of computation and HBM resources across $k_{\rm d}$ GPU devices. 
	In each decoding iteration, $N_{\rm bat}$ requests in a batch are decoded in parallel, each producing one output token at a time. 
	As new requests arrive or existing requests terminate after different numbers of iterations, requests are dynamically admitted to or removed from the batch, so the value of $N_{\rm bat}$ varies over time. 
	
	Let $L_{\rm io}$ denote the 
	total number of tokens 
	within the batch of user requests as
	\begin{equation}
		L_{\rm io} = \sum_{n=1}^{N_{\rm bat}} \left( L_{\rm i}^{(n)} + L_{\rm o}^{(n)} \right),\label{eq:LIO}
	\end{equation}
	where $L_{\rm i}^{(n)}$ and $L_{\rm o}^{(n)}$ denote the number of input tokens and the number of 
	output tokens already generated for the $n$-th user request, respectively.
	The required number of multiplications is then given by
	\begin{align}
			&{N}_{\rm mul,d}(L_{\rm io})\label{eq:dec}\\
			={}&N_{\rm layer}\left( N_{\rm bat} \left( {2(1+g) N_{\rm mod}^2}\!+\!{2 N_{\rm mod}N_{\rm hid}} \right)\!+\!2 N_{\rm mod} L_{\rm io}
			\right).\notag
	\end{align}
	Therefore, the decode computational latency is modeled as 
	\begin{equation}
		T_{\rm d,1}(L_{\rm io}) = \frac{N_{\rm mul,d}(L_{\rm io}) }{k_{\rm d} F_{\rm d}},\label{eq:Td1}
	\end{equation} 
	where $F_{\rm d}$ denotes the computational capacity of each decoding GPU.

	In each decoding iteration, all KV caches of the requests in a batch must be loaded into the computing units. Because of the large memory footprint of KV cache $g\gamma L_{\rm io}$ within the batch, the resulting latency
	\begin{equation}
		T_{\rm d,2}(L_{\rm io}) = \frac{\Omega_{\rm LLM}+ \displaystyle g\gamma L_{\rm io}
		}{k_{\rm d} B_{\rm HBM}}\label{eq:Td2}
	\end{equation}
	is non-negligible. Assuming perfect parallelism between computation and HBM read/write operations, the overall decode service time 
	can be written as
	\begin{equation}
		T_{\rm d}(L_{\rm io}) = \max\left( T_{\rm d,1},T_{\rm d,2} \right).%
		\label{eq:decodetime}
	\end{equation}

	\subsection{Stage-Wise LLM Service Metrics}\label{sec:qosmetric}
	To evaluate the SLO of LLM services, the following metrics are commonly considered.
	\begin{itemize}
		\item {\bf Time to First Token (TTFT)}: A critical latency metric in the prefill stage. TTFT is defined as the elapsed time from the arrival of a user request to the generation of the very first output token. It can be expressed by the sojourn time of the prefill queuing model as
		\begin{equation}
			T_{\rm TTFT} = T_{\rm p}^{\rm W} + T_{\rm p}\left( L_{\rm i}\right),\label{eq:TTFT}
		\end{equation}
		where $T_{\rm p}^{\rm W}$ is the waiting time of the user request in the prefill stage. 
		TTFT is affected by the number of input tokens $L_{\rm i}$, model footprint $\Omega_{\rm LLM}$, and hardware compute capacity $F_{\rm p}$. 
		In real-time chatting or agentic workflows, minimizing TTFT is of great importance, as it directly reflects the user's waiting time. 
		\item {\bf KV-Cache Transfer Latency}: After the prefill stage, the generated KV cache and the first token ID are transferred to the decode resource pool.  
		The transfer latency consists of the waiting time $T_{\rm TR}^{\rm W}$ in the transfer queue and the actual transmission time $T_{\rm TR}$ given by
		\begin{equation}
			T_{\rm KV} = T_{\rm TR}^{\rm W} + T_{\rm TR}(L_{\rm i}).\label{eq:kvlatency}
		\end{equation}
		Although it is not commonly treated as a conventional SLO metric, it connects the prefill and decode stages and influences the overall service quality. We hence treat it as an independent stage-wise SLO metric.
		\item {\bf Time per Output Token (TPOT)}: A fundamental latency metric in the decode stage. TPOT is defined by the time required to generate each output token as
		\begin{equation}
			T_{\rm TPOT} = T_{\rm d}(L_{\rm io}),
		\end{equation}
		which is also a random variable affected by the context length $L_{\rm io}$, batch size $N_{\rm bat}$, and other system parameters. 
		Lower TPOT produces smoother and more responsive text generation, and a 
		suitable TPOT should not exceed the time interval in which users can comfortably read.
		
	\end{itemize}

	\subsection{Problem Formulation}
	In this subsection, we formulate the SLO-aware resource allocation problem for the disaggregated LLM serving system described in the previous subsections.%
	
	An SLO specifies a latency threshold that the related metric must remain below with a prescribed probability $p_{\rm SLO}$. Goodput denotes the user request throughput $\lambda$ adhering to the latency SLO targets, i.e., ensuring that $p_{\rm SLO}$ percent of the requests satisfy the designated latency requirements~\cite{wang2026taiji,distserve}. %
	In this work, we aim to maximize the system goodput by jointly allocating the overall cost. 
	For the aforementioned three inference stages, let ${\bf r} = \{ k_{\rm p}, B, k_{\rm d},N_{\rm bat} \}$ be the set of resources to be allocated, 
	the goodput maximization problem is then defined by
	\begin{subequations}
		\begin{align}
			\!\!\mathcal{P}_1:~~
			\underset{{\bf r} }{\max}~~
			&\lambda\\
			{\rm s.t.}~~
			&k_{\rm p} C_{\rm p}+B C_{\rm KV}+k_{\rm d}C_{\rm d}\le C_{\max},\label{eq:cost}\\
			&\!\Pr\left( T_{\rm TTFT}\leq \tau_{\rm pre} \right)\geq p_{\rm SLO}
			,\label{eq:slo1}\\
			&\!\Pr\left( T_{\rm KV}\leq \tau_{\rm KV} \right)\geq p_{\rm SLO}
			,\label{eq:slo2}\\
			&\!\Pr\left( T_{\rm TPOT}\leq \tau_{\rm dec} \right)\geq p_{\rm SLO}
			,\label{eq:slo3}\\
			&\rho_{\rm p}<1,~ \rho_{\rm KV}<1,~k_{\rm p},k_{\rm d},N_{\rm bat}\in\mathbb Z.
		\end{align}
	\end{subequations}
	
	In problem $\mathcal{P}_1$,~\eqref{eq:cost} denotes the cost constraint, where $C_{\rm p}$, $C_{\rm KV}$, and $C_{\rm d}$ are the unit costs of prefill instance, KV cache transfer bandwidth, and decode device, respectively, and $C_{\rm max}$ is the maximum cost to be allocated.~\eqref{eq:slo1}-\eqref{eq:slo3} are the SLO targets for prefill, KV cache transfer, and decode stages, respectively, where $\tau_{\rm pre}$, $\tau_{\rm KV}$, and $\tau_{\rm dec} $ denote the latency thresholds for $T_{\rm TTFT}$, $T_{\rm KV}$, and $T_{\rm TPOT}$, respectively.
		
	The probabilities in~\eqref{eq:slo1}-\eqref{eq:slo3} typically do not admit closed-form expressions due to the complicated behaviors and heterogeneous stochastic mechanisms. Therefore, in the following sections, we will analyze the stage-wise latency distributions, derive tractable approximations for the SLO-attainment probabilities, and establish the corresponding resource requirements.

	\section{Analysis}\label{sec:ana}
	In this section, we analyze the stage-wise $p_{\rm SLO}$-quantile latency tail of SLO constraints in $\mathcal P_1$, and derive the corresponding resource requirements with respect to corresponding decision variables.
	Under practical setups\footnote{\label{fn:1} Unless otherwise stated, examples of practical system parameters in this paper refer to NVIDIA A100 GPU~\cite{a100} with $F_{\rm p} = 156\times 10^{12}$ multiplications per second and $B_{\rm HBM} = 2$ TBps, LLaMA 3.1 8B LLM~\cite{grattafiori2024llama3herdmodels} with $N_{\rm layer} = 32$, $N_{\rm mod} = 4,096$, $N_{\rm hid} = 14,336$, $g=0.25$, and $L_{\rm i} = 1,024$.}, the prefill and decode stages are bottlenecked by computational capacity and memory bandwidth under practical setups, respectively (See Appendix~A for the proof). 
	Therefore, we only consider latency arising from compute capacity in the prefill stage and memory bandwidth in the decode stage. 
	
		\subsection{Prefill Stage}
	\label{sec:prefillanalysis}
	
	In this subsection, we derive the closed-form approximation of sojourn latency $T_{\rm TTFT} = T_{\rm p}^{\rm W} + T_{\rm p}\left( L_{\rm i}\right)$ in the prefill stage. 
	
	We first derive the behavior of the prefill waiting time $T_{\rm p}^{\rm W}$. 
	With a limited overall cost $C_{\rm max}$, the prefill queue can operate at very high utilization, i.e., $\rho_{\rm p}\to 1$. In this case, the Kingman's approximation~\cite{kingman1992poisson,whitt2002stochastic} can be applied to the waiting time $T_{\rm p}^{\rm W}$ of M/G/$k_{\rm p}$ queue to analyze the tail behavior as
	\begin{equation}
		\Pr\left( T_{\rm p}^{\rm W}\geq \tau \right) \approx C_{k_{\rm p}}\!\!\left( A \right) e^{-\frac{\tau}{w}},\label{eq:prefillapprox}
	\end{equation}
	where 
	\begin{equation}
		C_{k_{\rm p}}\!\!\left( A \right)
		=
		\frac{\dfrac{A^{k_{\rm p}}}{k_{\rm p}!}\dfrac{1}{1-\rho_{\rm p}}}
		{\displaystyle
			\sum_{n=0}^{k_{\rm p}-1}\frac{A^n}{n!}
			+\dfrac{A^{k_{\rm p}}}{k_{\rm p}!}\dfrac{1}{1-\rho_{\rm p}}}
	\end{equation}
	is the Erlang-C waiting probability with traffic intensity $A = \rho_{\rm p}k_{\rm p}$,  %
	\begin{equation}
		w = \frac{1+c^2}{2} \frac{\rho_{\rm p}}{\lambda\left( 1-\rho_{\rm p} \right)}%
	\end{equation}
	is the approximation factor, 
	and $c = \sqrt{{\rm Var}[L_{\rm i}]}/\mathbb{E}[L_{\rm i}]$ is the coefficient of variation (CV) of random variable $L_{\rm i}$. 
	
	Next, we derive effective approximations for the service time in the prefill stage $T_{\rm p}\left( L_{\rm i}\right)$. 
	According to Appendix~A, we have $T_{\rm p}\approx T_{\rm p,1} = N_{\rm mul}( L_{\rm i} )/F_{\rm p}$. Under practical parameters\footref{fn:1}, the quadratic term in $N_{\rm mul}( L_{\rm i} )$ shows a limited contribution as 
	\begin{equation}
		\frac{N_{\rm layer} N_{\rm mod}L_{\rm i}^2}{N_{\rm mul}(L_{\rm i})}
		= 0.0256, %
	\end{equation}
	which indicates that the quadratic term in $N_{\rm mul}(L_{\rm i})$ has a limited effect (below $2.6\%$)  on the overall computational complexity. Therefore, the prefill service time can be further approximated as
	\begin{equation}
		T_{\rm p}
		\approx \alpha_{\rm p} L_{\rm i},\label{eq:servicetimeapprox}
	\end{equation}
	where $\alpha_{\rm p} = N_{\rm layer}( 2(1+g)N_{\rm mod}^2\!+\!(2N_{\rm hid}+1) N_{\rm mod} )/F_{\rm p}$.

	Since the waiting time $T_{\rm p}^{\rm W}$ and the service time $T_{\rm p}(L_{\rm i})$ are independent, the CDF $F_{\rm TTFT}(\tau_{\rm pre})$ of sojourn time $T_{\rm TTFT}$ is obtained by the convolution as
	\begin{equation}
		\begin{aligned}
			&\!\Pr\left( T_{\rm TTFT}\!\leq\! \tau_{\rm pre} \right)\\
			={}& \int_{0}^{\tau_{\rm pre}}\!\! \Pr\left( T_{\rm p}^{\rm W}\leq \tau_{\rm pre} - s \right)
			{\rm d} F_{T_{\rm p}}( s )
			\\
			\approx{}&\int_{0}^{\tau_{\rm pre}}\!\!\! \left( 1 \!-\! C_{k_{\rm p}}\!\!\left( A \right) e^{-\frac{\tau_{\rm pre}-s}{w}} \right)\!{\rm d} F_{T_{\rm p}}( s ) \\
			={}&F_{T_{\rm p}}( \tau_{\rm pre} ) - C_{k_{\rm p}}\!\!\left( A \right)\int_{0}^{\tau_{\rm pre}} \!\! e^{-\frac{\tau_{\rm pre}-s}{w}} {\rm d} F_{T_{\rm p}}( s ),
		\end{aligned}
	\end{equation}
	where $F_{T_{\rm p}}(\cdot)$ is the cumulative distribution function (CDF) of prefill service time $T_{\rm p}$. 
	The latency SLO constraint~\eqref{eq:slo1} is then reformulated as
	\begin{equation}
		F_{T_{\rm p}}( \tau_{\rm pre} ) - C_{k_{\rm p}}\!\left( A \right)\int_{0}^{\tau_{\rm pre}} \! e^{-\frac{\tau_{\rm pre}-s}{w}} {\rm d} F_{T_{\rm p}}( s ) \geq  p_{\rm SLO}.
		\label{eq:qoslhs1}
	\end{equation}
	In particular,~\eqref{eq:qoslhs1} provides a unified characterization for any given distribution of $L_{\rm i}$. 
	Based on 
	application scenarios, the distribution\footnote{While $L_{\rm i}$ is a discrete random variable in practice, to facilitate analysis, we model its distribution in the continuous domain. Similar treatment applies to other random variables such as $L_{\rm o}$ and $L_{\rm io}$ in the remainder of this paper.} of $L_{\rm i}$ can be categorized into two types~\cite{316730}:
	\begin{itemize}
		\item {\bf Light-tailed}: Short interactive workloads, such as conversation and code completion, are commonly dominated by relatively short prompts and can be modeled by an exponential distribution~\cite{patel2024splitwise,10946802}, i.e., $L_{\rm i}\sim {\rm Exp}[1/\ell_{\rm i}]$, where $\ell_{\rm i} = \mathbb{E}[L_{\rm i}]$.  In this case,~\eqref{eq:prefillapprox} is an exact equality instead of an approximation.%
		\item {\bf Heavy-tailed}: Long context workloads, such as multi-modal and agentic tasks with multi-turn dialogues, usually exhibit a heavy-tailed length distribution that can be modeled by a log-normal distribution~\cite{ModServe}, i.e., $L_{\rm i}\sim \mathcal{LN}[\mu_{\rm i}, \sigma_{\rm i}^2]$.%
	\end{itemize}
	For light-tailed case with $L_{\rm i}\sim {\rm Exp}[1/\ell_{\rm i}]$, the CDF of $T_{\rm p}(L_{\rm i})$ is given by $F_{T_{\rm p}}(t) = 1-e^{-\frac{t}{\alpha_{\rm p} \ell_{\rm i}}}$. The integral form in~\eqref{eq:qoslhs1} for constraint~\eqref{eq:slo1} is then reduced to
	\begin{equation}
		1-e^{-\frac{\tau_{\rm pre}}{\alpha_{\rm p}\ell_{\rm i}}} - \frac{w C_{k_{\rm p}}(A)}{\alpha_{\rm p}\ell_{\rm i}-w}\left( e^{-\frac{\tau_{\rm pre}}{\alpha_{\rm p}\ell_{\rm i}}} -e^{-\frac{\tau_{\rm pre}}{w}} \right)\geq p_{\rm SLO}.
	\end{equation}
	However, for the heavy-tailed case with $L_{\rm i}\sim \mathcal{LN}[\mu_{\rm i}, \sigma_{\rm i}^2]$, the convolution integral~\eqref{eq:qoslhs1} does not admit a finite closed-form expression and should therefore be evaluated numerically. 
	Since an SLO constraint physically corresponds to a minimal resource requirement, the unique optimal $k_{\rm p}^\star$ can be obtained via enumeration. 
	
		\begin{figure}[t]
		\centering
		\includegraphics[width=0.4\textwidth]{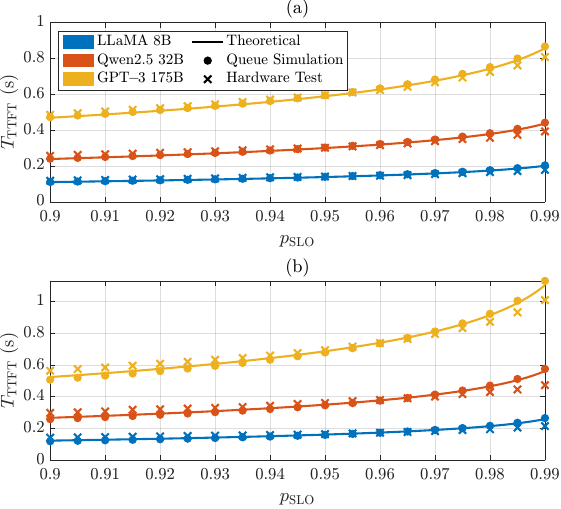}
		\caption{TTFT metric with varying $p_{\rm SLO}$ thresholds for (a) light-tailed and (b) heavy-tailed workloads, respectively.}
		\label{fig:val_pref}
	\end{figure}
	
	The approximations in~\eqref{eq:prefillapprox} and~\eqref{eq:servicetimeapprox} are validated in Fig.~\ref{fig:val_pref}. 
	In Fig.~\ref{fig:val_pref}(a), the only approximation stems from~\eqref{eq:servicetimeapprox} since~\eqref{eq:prefillapprox} is exact in the light-tailed case. In other words, we show that~\eqref{eq:servicetimeapprox} is an accurate approximation. 
	In Fig.~\ref{fig:val_pref}(b), 
	the average deviation over the entire range of $p_{\rm SLO}$ is within $5\%$, which validates the accuracy of the closed‑form approximations in~\eqref{eq:prefillapprox} and~\eqref{eq:servicetimeapprox} for the heavy-tailed case. 
	{In addition, the hardware measurements closely follow the theoretical predictions. The average relative error remains below $5\%$ for $p_{\rm SLO}<0.98$, and the discrepancy becomes more visible only at the extreme tail percentiles.}

	\subsection{KV Cache Transfer Stage}
	In this subsection, we analyze the sojourn latency $T_{\rm KV}$ in the KV cache transfer stage, and derive approximate closed-form solutions for the light-tailed and heavy-tailed cases, respectively. 
	
	In the KV cache transfer stage, the communication link with bandwidth $B$ serves one KV cache transfer request. Therefore, the KV cache transfer stage is modeled as an M/G/$1$ queue. For user requests with light- and heavy-tailed characteristics, the sojourn latencies are analyzed as follows.
	
	\subsubsection{\bf Light-Tailed}
	For the number of tokens following a light-tailed exponential distribution as $L_{\rm i}\sim {\rm Exp}[1/\ell_{\rm i}]$, the service time $T_{\rm TR} = g\gamma L_{\rm i}/B$ is also exponentially distributed. Therefore, the queuing model of the KV cache transfer stage degenerates to an M/M/$1$ queue. 
	In this case, the waiting time $T_{\rm TR}^{\rm W}$ also follows an exponential distribution with probability mass $1-\rho_{\rm KV}$ at $t=0$. Therefore, the CDF 
	of the waiting time is given by~\cite{90319}
	\begin{equation}
		\Pr\left(T_{\rm TR}^{\rm W}\leq t\right)= 1-\rho_{\rm KV} e^{-\frac{t}{w_{\rm KV}}},
		\label{eq:kv_waiting_lt}
	\end{equation}
	where 
	\begin{equation}
		w_{\rm KV} = \frac{\rho_{\rm KV}}{\lambda\left( 1-\rho_{\rm KV}\right)}
	\end{equation}
	is the tail decay factor. Since the waiting time and the service time of an arriving request are independent, the CDF of the sojourn latency $T_{\rm KV}$ can be obtained by the convolution as
	\begin{align}
		\!\Pr\left( T_{\rm KV}\!\leq\! \tau_{\rm KV} \right)
		={}&\!\!\int_{0}^{\tau_{\rm KV}}\!\! \Pr\left( T_{\rm TR}^{\rm W}\leq \tau_{\rm KV} - t \right)  
		{\rm d} \Pr\left( T_{\rm TR}\leq  t \right) \notag\\
		={}&1-e^{-\frac{\tau_{\rm KV}}{w_{\rm KV}}}.
	\end{align}
	Therefore, the SLO constraint~\eqref{eq:slo2} is exactly characterized as
	\begin{equation}
		1-e^{-\frac{\tau_{\rm KV}}{w_{\rm KV}}}\geq p_{\rm SLO},
	\end{equation}
	which yields
	\begin{equation}
		B\geq \lambda g\gamma \ell_{\rm i} + \frac{g\gamma \ell_{\rm i}}{\tau_{\rm KV}}\log\frac{1}{1-p_{\rm SLO}}.\label{eq:kvlight}
	\end{equation}

	\subsubsection{\bf Heavy-Tailed}
	For multi-modal and agentic requests, the input length $L_{\rm i}$ is modeled by a log-normal distribution as $L_{\rm i}\sim\mathcal{LN}[\mu_{\rm i},\sigma_{\rm i}^{2}]$, where $\mu_{\rm i}$ and $\sigma_{\rm i}$ are the corresponding distribution parameters. 
	According to~\eqref{eq:TKV}, the transfer service time also follows a log-normal distribution as
	\begin{equation}
		T_{\rm TR}(L_{\rm i})
		=\frac{g\gamma L_{\rm i}}{B}
		\sim
		\mathcal{LN}\left[
		\mu_{\rm i}+\log\left(\frac{g\gamma}{B}\right),
		\sigma_{\rm i}^{2}
		\right].
	\end{equation}
	Unfortunately, with heavy-tailed service time behavior, the waiting time distribution of the M/G/$1$ queue does not admit the tractable exact expression available in the light-tailed case. 
	Hence, we resort to the Kingman-type heavy traffic approximation~\cite{kingman1992poisson,whitt2002stochastic} similar to the prefill stage and obtain
	\begin{equation}
		B\geq \lambda g\gamma \ell_{\rm i} + \frac{e^{\sigma_{\rm i}^2}}{2}\frac{g\gamma \ell_{\rm i}}{\tau_{\rm KV}}\log\frac{\rho_{\rm KV}}{1-p_{\rm SLO}}.%
		\label{eq:kvheavy}
	\end{equation}
	For detailed derivation, please refer to Appendix~B. 
	Hence, we can obtain optimal $B^\star$ via one-dimensional line search.
	
	\begin{figure}[t]
		\centering
		\includegraphics[width=0.4\textwidth]{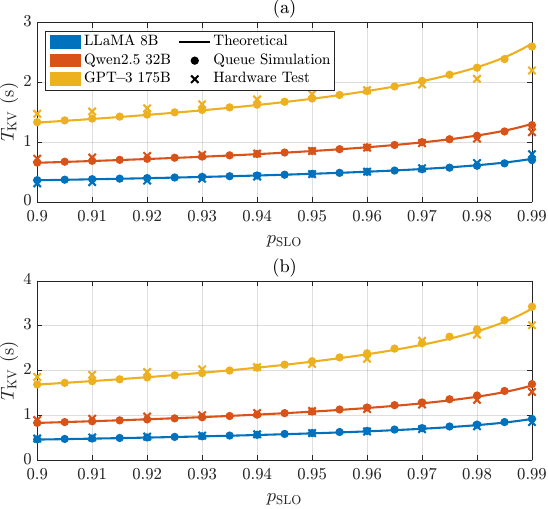}
		\caption{KV cache transfer latency with varying $p_{\rm SLO}$ thresholds for (a) light-tailed and (b) heavy-tailed workloads, respectively.}
		\label{fig:val_kv}
	\end{figure}
	
	Figs.~\ref{fig:val_kv}(a) and~\ref{fig:val_kv}(b) validate the accuracy of the closed-form expressions~\eqref{eq:kvlight} and~\eqref{eq:kvheavy} for the light- and heavy-tailed cases, respectively. In particular, the theoretical predictions match the queue simulations and hardware measurement results with negligible errors. The deviations increase only at the extreme high tails, where system-level overheads that are not considered in the queuing model may have a stronger impact.
	
	\begin{remark} 
		It is worth noting that for both light- and heavy-tailed workloads, the bandwidth $B$ requirements on the right-hand-side of~\eqref{eq:kvlight} and~\eqref{eq:kvheavy} can be divided into a $\lambda$-related term and an SLO-related term, while the SLO-related 
		bandwidth requirement is independent of the request arrival rate $\lambda$. This indicates that when $\lambda$ changes, we only need to scale the $\lambda$-related bandwidth linearly with the factor $g\gamma \ell_{\rm i}$ to meet the SLO target.
	\end{remark}

		\subsection{Decode Stage}

	In this subsection, we analyze the TPOT constraint~\eqref{eq:slo3} in $\mathcal{P}_1$. Different from the prefill and KV cache transfer stages, the resource allocation in the decode stage is determined by two variables, i.e., the number of decode GPU devices $k_{\rm d}$ and the maximum batch size $N_{\rm bat}$. Increasing $N_{\rm bat}$ improves the admission capability of user requests, but also increases the requirements of HBM capacity $k_{\rm d}\Omega_{\rm HBM}$ and total bandwidth $k_{\rm d}B_{\rm HBM}$. In the following, we derive the relationships between $k_{\rm d}$ and $N_{\rm bat}$ to determine the feasible region for resource allocation.

	According to Appendix~A, the decode service time is mainly determined by the HBM bandwidth under practical configurations as 
	\begin{equation}
		T_{\rm TPOT}\approx T_{\rm d,2} = \frac{\Omega_{\rm LLM}+ \displaystyle g\gamma L_{\rm io}
		}{k_{\rm d} B_{\rm HBM}}.
		\label{eq:tpotapprox}
	\end{equation}
	where the $L_{\rm io}$ is the aggregated number of tokens within the batch defined by~\eqref{eq:LIO}. Different from the prefill stage, where $L_{\rm i}$ follows either a light-tailed exponential or heavy-tailed log-normal distribution, $L_{\rm o}$ statistically follows an exponential distribution as $L_{\rm o}\sim {\rm Exp}[1/\ell_{\rm o}]$, and is typically independent with $L_{\rm i}$~\cite{10946802,ModServe,316730}.

	According to~\eqref{eq:tpotapprox}, the $p_{\rm SLO}$-quantile of TPOT is closely related to that of $L_{\rm io}$. {Therefore, in the following, we analyze the TPOT SLO constraint under light- and heavy-tailed cases using moment matching~\cite{1097606,8094298}, respectively.}
	\begin{itemize}
		\item {\bf Light-tailed}: In this case, $L_{\rm io}$ is the sum of $N_{\rm bat}$ exponential random variables $L_{\rm i}\sim {\rm Exp}[1/\ell_{\rm i}]$ and $N_{\rm bat}$ exponential random variables $L_{\rm o}\sim {\rm Exp}[1/\ell_{\rm o}]$. Therefore, we can approximate the distribution of $L_{\rm io}$ by using a Gamma distribution random variable with a constant bias~\cite{8094298}. 
		
		Specifically, let $L_{\rm io}^\prime = \ell_{\rm LT} + \tilde{L}_{\rm io}$ be the approximated random variable of $L_{\rm io}$, where $\tilde{L}_{\rm io}\sim {\rm Gamma}[\alpha_{\rm LT}, \theta_{\rm LT}]$ is a Gamma-distributed random variable, and $\ell_{\rm LT}$ is a constant bias. 
		The parameters $\{ \ell_{\rm LT},\alpha_{\rm LT}, \theta_{\rm LT} \}$ can be obtained by matching the mean, variance, and skewness of $L_{\rm io}$ as
		\begin{equation}
			\begin{cases}
				\ell_{\rm LT} \!\!\!\!& =\displaystyle N_{\rm bat} \frac{\ell_{\rm i}\ell_{\rm o}\left(\ell_{\rm i}-\ell_{\rm o}\right)^2}{\ell_{\rm i}^3+\ell_{\rm o}^3}\\
				\alpha_{\rm LT} \!\!\!\!&=\displaystyle N_{\rm bat}\frac{\left(\ell_{\rm i}^2+\ell_{\rm o}^2\right)^3}{\left(\ell_{\rm i}^3+\ell_{\rm o}^3\right)^2}\\
				\theta_{\rm LT} \!\!\!\!&=\displaystyle \frac{\ell_{\rm i}^3+\ell_{\rm o}^3}{\ell_{\rm i}^2+\ell_{\rm o}^2}
			\end{cases}.
		\end{equation}
		Therefore, the $p_{\rm SLO}$-quantile of $L_{\rm io}^\prime$ is given by
		\begin{equation}
			\ell\left(p_{\rm SLO}\right) = \ell_{\rm LT} + \theta_{\rm LT} P^{-1}_{\alpha_{\rm LT}}(p_{\rm SLO}),
		\end{equation}
		where $P^{-1}_{\alpha_{\rm LT}}(\cdot)$ is the inverse function of the regularized lower incomplete gamma function with parameter $\alpha_{\rm LT}$.
			
		\item {\bf Heavy-tailed}: In this case, $L_{\rm io}$ is the sum of $N_{\rm bat}$ log-normal random variables $L_{\rm i}\sim \mathcal{LN}[\mu_{\rm i},\sigma_{\rm i}^2]$ and $N_{\rm bat}$ exponential random variables $L_{\rm o}\sim {\rm Exp}[1/\ell_{\rm o}]$. Therefore, we can model $L_{\rm io}$ using a log-normal random variable with a constant bias~\cite{1097606}. 
		
		Specifically, let $L_{\rm io}^\prime = \ell_{\rm HT} + \tilde{L}_{\rm io}$ be the approximated random variable of $L_{\rm io}$, where $\tilde{L}_{\rm io}\sim \mathcal{LN}[\mu_\ell, \sigma_\ell^2]$ is a log-normal random variable, and $\ell_{\rm HT}$ is a constant bias. 
		The parameters $\{ \ell_{\rm HT},\mu_\ell, \sigma_\ell \}$ can be obtained by matching the mean, variance, and skewness of $L_{\rm io}$ as
		\begin{equation}
			\begin{cases}
				\ell_{\rm HT} \!\!\!\!& =\displaystyle m - \frac{\sqrt{v}}{x}\\
				\mu_\ell \!\!\!\!&=\displaystyle \frac{1}{2}\log\frac{v}{1+x^2} -\log x\\
				\sigma_\ell^2 \!\!\!\!&=\displaystyle \log\left(1+x^2 \right)
			\end{cases},
		\end{equation}
		where 
		\begin{equation}
			\begin{cases}
				m \!\!\!\!&=\displaystyle N_{\rm bat} \left(\ell_{\rm o} + e^{\mu_{\rm i} + \sigma_{\rm i}^2/2}\right)\\
				v \!\!\!\!&=\displaystyle N_{\rm bat} \left(\ell_{\rm o}^2 + (e^{\sigma_{\rm i}^2}-1)e^{2\mu_{\rm i}+\sigma_{\rm i}^2}\right)\\
				x \!\!\!\!&=\displaystyle 2\sinh\left( \frac{1}{3}{\rm arcsinh}\left(\frac{s}{2}\right) \right)\\
				s \!\!\!\!&=\displaystyle \frac{e^{3\mu_{\rm i}+\frac{3}{2}\sigma_{\rm i}^2}
					\left(e^{\sigma_{\rm i}^2}+2\right)
					\left(e^{\sigma_{\rm i}^2}-1\right)^2
					+2\ell_{\rm o}^3}{\sqrt{N_{\rm bat}\left( e^{2\mu_{\rm i}+\sigma_{\rm i}^2}
						\left(e^{\sigma_{\rm i}^2}-1\right)+\ell_{\rm o}^2\right)^3}}
			\end{cases}.
		\end{equation}
		Therefore, the $p_{\rm SLO}$-quantile of $L_{\rm io}^\prime$ is given by
		\begin{equation}
			\ell( p_{\rm SLO} ) = \ell_{\rm HT} + e^{\mu_\ell + \sigma_\ell \Phi^{-1}\left( p_{\rm SLO} \right)},\label{eq:pquantLio}
		\end{equation}
		where $\Phi(\cdot)$ is the CDF of the standard normal distribution. 
	\end{itemize} 
	\noindent Therefore, the probability on the left-hand-side of~\eqref{eq:slo3} in $\mathcal{P}_1$ can be represented by $L_{\rm io}^\prime$ as
	\begin{equation}
		\begin{aligned}
			\!\!\Pr\!\left(T_{\rm TPOT} \!\leq\! {\tau_{\rm dec}} \right) \!\approx\! \Pr\!\left(\! L_{\rm io}^\prime \!\leq\! \frac{\tau_{\rm dec}k_{\rm d}B_{\rm HBM} - \Omega_{\rm LLM}}{g\gamma} \!\right).
			\label{eq:decodeqos}
		\end{aligned}
	\end{equation}
	Based on the workload characteristics, the SLO constraint~\eqref{eq:slo3} is equivalent to
	\begin{equation}
		{\rm C1\,(L)}:~~k_{\rm d}\geq \frac{ g\gamma  \left(  \ell_{\rm LT}+ \theta_{\rm LT} P^{-1}_{\alpha_{\rm LT}}(p_{\rm SLO}) \right) + \Omega_{\rm LLM} }{\tau_{\rm dec} B_{\rm HBM}}\label{eq:dec_ub1l}
	\end{equation}
	for the light-tailed workload, or
	\begin{equation}
		{\rm C1\,(H)}:~~k_{\rm d}\geq \frac{ g\gamma  \left(  \ell_{\rm HT}+e^{\mu_\ell + \sigma_\ell\Phi^{-1}(p_{\rm SLO})} \right) + \Omega_{\rm LLM} }{\tau_{\rm dec} B_{\rm HBM}}\label{eq:dec_ub1h}
	\end{equation}
	for the heavy-tailed workload. These conditions ensure that $p_{\rm SLO}$ percent of the requests satisfy the SLO~\eqref{eq:slo3}. They provide one of the boundaries for decision variables $\{ k_{\rm d}, N_{\rm bat} \}$. 
	
	Another boundary is given by the HBM capacity. Other than the model weights, each active user request needs to maintain the KV cache associated with its input tokens and currently generated output tokens. To ensure that the HBM capacity is not exhausted with probability $p_{\rm mem}$, the following inequality must hold as
	\begin{equation}
		\Omega_{\rm LLM}
		+ g\gamma \ell( p_{\rm mem} )
		\le k_{\rm d}\Omega_{\rm HBM},\label{eq:C2_cond}
	\end{equation}
	which yields
	\begin{equation}
		\!{\rm C2\,(L)}:~k_{\rm d}\geq \frac{ g\gamma  \left(  \ell_{\rm LT}+\theta_{\rm LT} P^{-1}_{\alpha_{\rm LT}}(p_{\rm mem}) \right) + \Omega_{\rm LLM} }{ \Omega_{\rm HBM}}\label{eq:dec_ub2l}
	\end{equation}
	or
	\begin{equation}
		\!{\rm C2\,(H)}:~k_{\rm d}\geq \frac{ g\gamma  \left(  \ell_{\rm HT}+e^{\mu_\ell + \sigma_\ell\Phi^{-1}(p_{\rm mem})} \right) + \Omega_{\rm LLM} }{ \Omega_{\rm HBM}}.\label{eq:dec_ub2h}
	\end{equation}

	Note that $\rm C1$ and $\rm C2$ both implicitly limit the batch size $N_{\rm bat}$ for a given $k_{\rm d}$, i.e., $\rm C1$ limits $N_{\rm bat}$ to satisfy the TPOT SLO constraint, while $\rm C2$ limits $N_{\rm bat}$ to prevent HBM capacity exhaustion. 
	Nevertheless, the batch size $N_{\rm bat}$ cannot be arbitrarily small. There are also two major constraints that contribute to the other two boundaries. 
	{The first one is the stability condition given by
	\begin{equation}
		\!\!{\rm C3}:~N_{\rm bat} >  \frac{\lambda \Omega_{\rm LLM}}{k_{\rm d}B_{\rm HBM} p_0 - \lambda g\gamma\left( \ell_{\rm i} +\ell_{\rm o} \right)}\label{eq:dec_lb1},%
	\end{equation}
	where $p_0$ denotes the probability of reaching the EOS token for each request. Since the output length $L_{\rm o}^{(n)}$ follows an exponential distribution as $L_{\rm o}^{(n)}\sim {\rm Exp}[1/\ell_{\rm o}]$, and is independent of the input length $L_{\rm i}^{(n)}$~\cite{10946802,316730,ModServe}, $p_0$ is further given by
	\begin{equation}
		p_0 = \Pr\left( L_{\rm o}\leq n+1\mid L_{\rm o}\geq n \right) = 1-e^{-\frac{1}{\ell_{\rm o}}}.
	\end{equation}
	Please refer to Appendix~C for more derivation details. 
	
	The second one is the direct admission condition given by
	\begin{equation} 
		{\rm C4}:~~N_{\rm bat}\geq \mu_{\rm bat} + \sigma_{\rm bat}\Phi^{-1}(p_{\rm join}),
		\label{eq:dec_lb2}
	\end{equation}
	where $\mu_{\rm bat}$ and $\sigma_{\rm bat}$ are the mean and standard deviation of the number of requests in a batch, and $p_{\rm join}$ is the probability that a request can directly join the decoding batch without waiting. Please refer to Appendix~D for more details.} 
			
	$\rm C4$ is typically a tighter condition than $\rm C3$, since it intrinsically implies the stability of the decoding process, and requires a larger batch size $N_{\rm bat}$ to reduce the waiting time.
	
	\begin{figure*}[t]
		\centering
		\includegraphics[width=\textwidth]{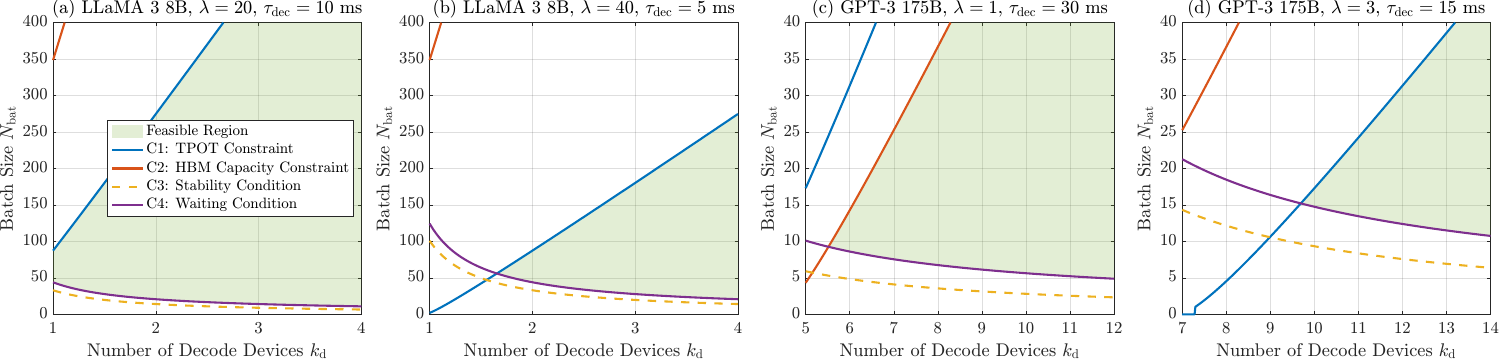}
		\caption{The heavy-tailed decision boundaries $\rm C1 - C4$ and the corresponding feasible regions for (a)-(b) LLaMA 3.1 8B and (c)-(d) GPT-3 175B models.}
		\label{fig:fea}
	\end{figure*}
	
	\begin{figure}[t]
		\centering
		\includegraphics[width=0.4\textwidth]{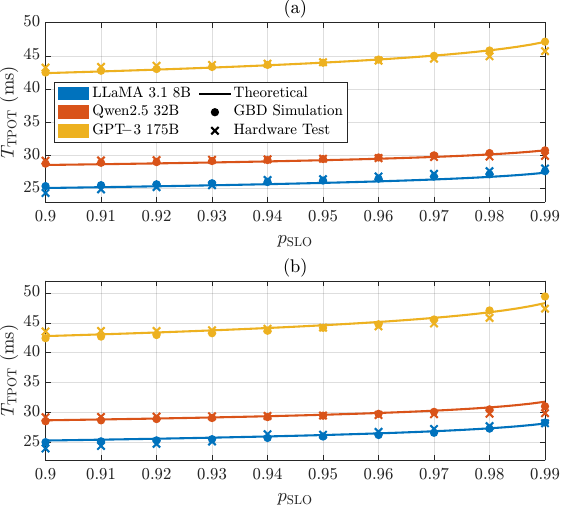}
		\caption{TPOT with varying $p_{\rm SLO}$ thresholds for (a) light-tailed and (b) heavy-tailed workloads, respectively.}
		\label{fig:val_dec}
	\end{figure}
	
	Fig.~\ref{fig:fea} depicts the feasible regions delineated by boundaries $\rm C1-C4$ for LLaMA 3.1 and GPT-3 models under varying system parameters and heavy-tail user request distributions. For a fixed $k_{\rm d}$, $\rm C1$ and $\rm C2$ act as upper constraints of batch size $N_{\rm bat}$, while $\rm C3$ and $\rm C4$ serve as lower constraints. 
	
	{Fig.~\ref{fig:val_dec} validates the proposed TPOT quantile characterization under both light- and heavy-tailed workloads. The analytical predictions closely match the simulation results. Meanwhile, the deviation between the analytical results and hardware measurements falls below $5\%$ over most of the considered $p_{\rm SLO}$ range. These results demonstrate that the proposed model accurately captures the decoding latency behavior under practical hardware execution.}
	
	\begin{algorithm}[t]
		\caption{Proposed Resource Allocation Algorithm}\label{alg:opt}
		\begin{algorithmic}[1]
			\REQUIRE The cost budget $C_{\max}$, the prescribed SLO thresholds $\{ \tau_{\rm pre}, \tau_{\rm KV}, \tau_{\rm dec} \}$ and $p_{\rm SLO}$, the decode reliability parameters $p_{\rm mem}$ and $p_{\rm join}$ for $\rm C2$ and $\rm C4$, the unit costs $C_{\rm p}$, $C_{\rm KV}$, and $C_{\rm d}$, and the parameters of model and hardware
			\ENSURE The maximum goodput $\lambda ^\star$
			\STATE Initialize trial solution $\lambda ^{(0)}$ and iteration index $i = 0$
			\REPEAT[$i\leftarrow i+1$]
			\STATE Compute the minimum prefill instance requirement $k_{\rm p}(\lambda^{(i)})$ via~\eqref{eq:qoslhs1}
			\STATE Compute the minimum KV transfer bandwidth requirement $B(\lambda^{(i)})$ via~\eqref{eq:kvlight} or~\eqref{eq:kvheavy}
			\STATE Compute the minimum decode device requirement $k_{\rm d}(\lambda^{(i)},N_{\rm bat})$ via ${\rm C1}-{\rm C4}$
			\STATE Compute the overall cost 
			\begin{equation}
				C^{(i)} = k_{\rm p}(\lambda^{(i)})C_{\rm p}+B(\lambda^{(i)})C_{\rm KV}+k_{\rm d}(\lambda^{(i)},N_{\rm bat})C_{\rm d}\notag
			\end{equation}
			\STATE Bisectionally update $\lambda ^{(i)}$ by comparing $C^{(i)}$ and $C_{\rm max}$
			\UNTIL $\vert \lambda ^{i} - \lambda ^{i-1} \vert \leq \varepsilon$
			\RETURN Maximum goodput $\lambda ^\star = \lambda ^{i}$
		\end{algorithmic}
	\end{algorithm}
	
	\section{SLO-Aware Resource Allocation}

	In this section, we develop an SLO-aware resource allocation scheme for solving the goodput maximization problem $\mathcal P_1$. 
	
	According to the analysis in Sec.~\ref{sec:ana}, the minimum required resources ${\bf r} = \{ k_{\rm p}, B, k_{\rm d}, N_{\rm bat} \}$ can be obtained given the arrival rate $\lambda$ as well as SLOs, which yield the minimum deployment cost $C_{\rm min}$. Since $C_{\rm min}$ increases monotonically with arrival rate $\lambda$, we can then obtain the maximum goodput via a bisection search. 
	Therefore, we can solve the following cost minimization problem as
	\begin{subequations}
		\begin{align}
			\!\!\mathcal{P}_2:~~
			\underset{{\bf r} }{\max}~~
			&\lambda\\
			{\rm s.t.}~~
			&k_{\rm p} C_{\rm p}+B C_{\rm KV}+k_{\rm d}C_{\rm d}\le C_{\max},\label{eq:cost2}\\
			&\eqref{eq:qoslhs1}
			,\\
			&\eqref{eq:kvlight}~{\rm or}~\eqref{eq:kvheavy}\label{eq:lightorheavy}\\
			&{\rm C1}-{\rm C4},
			,\label{eq:slo3_2}\\
			&\rho_{\rm p}<1,~ \rho_{\rm KV}<1,~k_{\rm p},k_{\rm d},N_{\rm bat}\in\mathbb Z.
		\end{align}
	\end{subequations}
		where~\eqref{eq:lightorheavy} is determined by the light-tailed or heavy-tailed workload type. The procedure of resource allocation is summarized in \textbf{Algorithm}~\ref{alg:opt}.

	\section{Insights}
	In this section, we revisit the analytical mappings established above and summarize the insights that are most relevant to the resource planning of disaggregated LLM serving systems.
	
	\begin{figure*}[t]
		\centering
		\includegraphics[width=\textwidth]{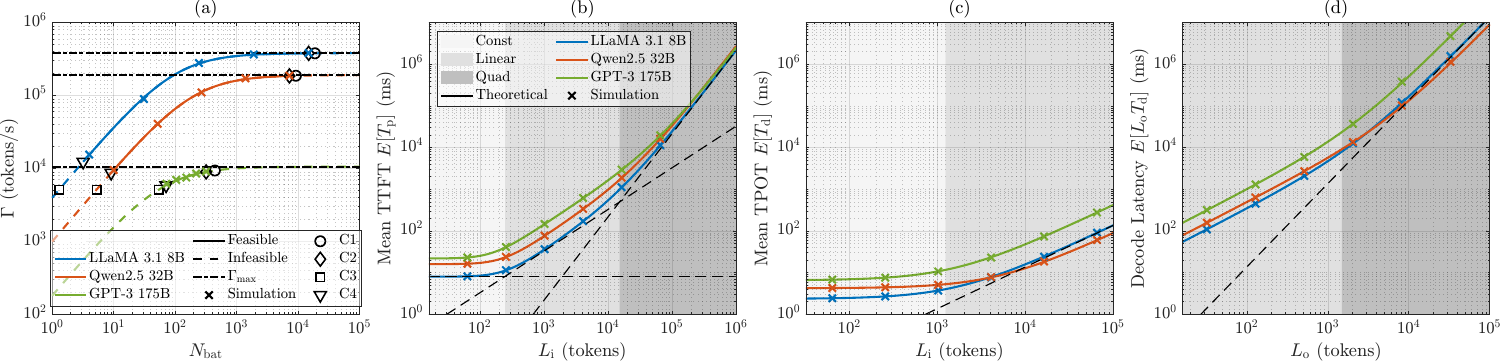}
		\caption{(a) Token throughput over varying batch size. (b)-(d) The inference latency metrics over varying lengths of input user requests $L_{\rm i}$ and output sequences $L_{\rm o}$.}
		\label{fig:insight1}
	\end{figure*}
	
	\subsection{Token Throughput}
	For LLM serving systems, the token generation rate, i.e., the token throughput, is another key performance metric that measures the total serving capacity of the system. Under the proposed analytical framework, each batched request generates exactly one token per decoding iteration, whose duration is dominated by the HBM access overhead in~\eqref{eq:tpotapprox}. 
	Therefore, the token throughput is given by
	\begin{equation}
		\Gamma = \frac{N_{\rm bat}}{\mathbb{E}\left[T_{\rm TPOT}\right]} = \frac{k_{\rm d}B_{\rm HBM}N_{\rm bat}}{\Omega_{\rm LLM} + g\gamma N_{\rm bat}\left(\ell_{\rm i}+\ell_{\rm o}\right)},
		\label{eq:tput}
	\end{equation}
	where $N_{\rm bat}$ denotes the number of concurrent requests within the decoding batch. 
	Since $\partial \Gamma/\partial N_{\rm bat}> 0$ always holds, the token throughput always increases with the number of concurrent requests $N_{\rm bat}$. However, the fact that $\partial^2 \Gamma/\partial N_{\rm bat}^2< 0$ indicates that the throughput gain diminishes as the concurrency $N_{\rm bat}$ grows. 
	As $N_{\rm bat}$ continues to increase, the token throughput $\Gamma$ saturates at
	\begin{equation}
		\Gamma_{\max} = \lim_{N_{\rm bat}\to\infty}\Gamma = \frac{k_{\rm d}B_{\rm HBM}}{g\gamma\left(\ell_{\rm i}+\ell_{\rm o}\right)},
		\label{eq:tputmax}
	\end{equation}
	which depends only on the overall HBM bandwidth $k_{\rm d}B_{\rm HBM}$ of the decode pool and the KV cache footprint $\Omega_{\rm KV} = g\gamma(\ell_{\rm i}+\ell_{\rm o})$. 
	As shown in Fig.~\ref{fig:insight1}(a), under identical hardware configurations, LLaMA 3.1 8B and Qwen2.5 32B achieve higher saturation throughputs $\Gamma_{\max}$ than the GPT-3 175B, owing to their smaller per-token KV cache footprints and the use of GQA, characterized by $\gamma$ and $g$, respectively. 
	It is also worth noting that in the feasible region delineated by ${\rm C1}-{\rm C4}$, all models can achieve over $90\%$ of the maximum token throughput.

	\subsection{Quadratic Impact of Input and Output Lengths}

	With the popularization of agentic LLM workflows, scenarios with extremely long input user requests from multi-turn LLM conversations are becoming increasingly common. The expanding context lengths for both input $L_{\rm i}$ and output $L_{\rm o}$ impose distinct SLO pressures across different stages.
	
	First, as shown in Fig.~\ref{fig:insight1}(b), the TTFT latency versus $L_{\rm i}$ across different LLM models falls into three regions. 
	For small $L_{\rm i}$, i.e., $L_{\rm i} < \tilde{L}_{\rm i}$, TTFT is dominated by HBM bandwidth and thus remains largely constant. As $L_{\rm i}$ grows, TTFT scales approximately linearly. For sufficiently large $L_{\rm i}$, TTFT will increase quadratically with $L_{\rm i}$. 
	This is because the increase of $L_{\rm i}$ amplifies the impact of the quadratic term $L_{\rm i}^2$ in~\eqref{eq:nmul} during the prefill stage. 
	Under the stability condition $\rho_{\rm p}= \lambda \mathbb{E}[T_{\rm p}(L_{\rm i})]/k_{\rm p}<1$, we have $k_{\rm p} > \lambda \mathbb{E}[T_{\rm p}(L_{\rm i})]$, which suggests that, in multi-modal or agentic multi-turn dialogue scenarios, the requirement of prefill devices $k_{\rm p}$ also scales quadratically with the input length $L_{\rm i}$. 
	
	\begin{figure*}[t]
		\centering
		\includegraphics[width=0.75\textwidth]{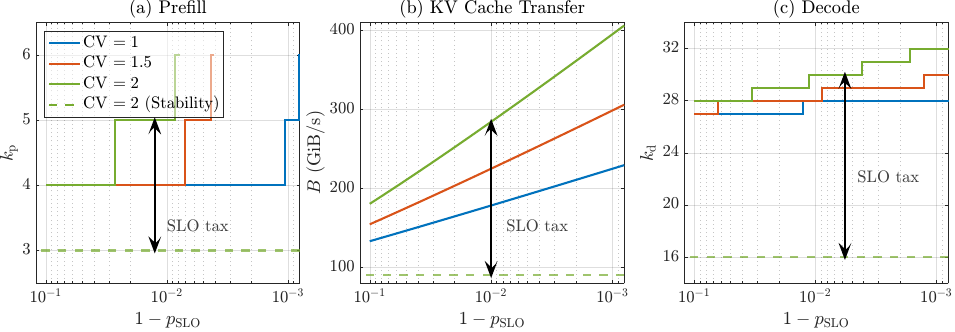}
		\caption{The resource requirements under varying SLO violation probability $1-p_{\rm SLO}$ and stability condition.}
		\label{fig:insight2}
	\end{figure*}
	
	Besides, both the input length $L_{\rm i}$ and output length $L_{\rm o}$ show a significant impact on the decode stage. Specifically, when the input (or output) sequence length $L_{\rm i}$ (or $L_{\rm o}$) is small, the LLM weight term $\Omega_{\rm LLM}$ outweighs KV cache term $g\gamma L_{\rm io}$ in~\eqref{eq:Td2}. 
	However, as either $L_{\rm i}$ or $L_{\rm o}$ increases, TPOT will eventually exhibit linear growth with $L_{\rm i}$ or $L_{\rm o}$, as illustrated in Fig.~\ref{fig:insight1}(c).

	Furthermore, 
	long output lengths $L_{\rm o}$ in complex tasks lead to an approximately quadratic effect on total decode time, as shown in Fig.~\ref{fig:insight1}(d). 
	When $\ell_{\rm o}$ is sufficiently large, the probability of reaching EOS token is approximated by $p_{0} = 1-e^{-\frac{1}{\ell_{\rm o}}} \approx {1}/{\ell_{\rm o}}$.
	According to the stability condition $\rm C3$, the requirement of decoding devices is given by
	\begin{equation}
		k_{\rm d}\gtrsim\frac{\lambda\left( \ell_{\rm o}\Omega_{\rm LLM} + g\gamma N_{\rm bat} \left( \ell_{\rm i}\ell_{\rm o}+\ell_{\rm o}^2 \right)\right)}{N_{\rm bat} B_{\rm HBM} }\sim\mathcal{O}(\ell_{\rm o}^2),
	\end{equation}
	indicating that $k_{\rm d}$ scales quadratically with the output length when $\ell_{\rm o}$ is large. 
	
	It is worth noting that these hardware requirements are derived from the stability condition, which are conservative estimates of the numbers of the prefill and decode devices required for stable operation. 
	Satisfying more stringent SLOs typically requires additional hardware capacity and consequently a substantially larger device allocation.

	\subsection{The SLO Tax}

	Compared to the stability conditions, i.e., $\rho_{\rm p}<1$, $\rho_{\rm KV}<1$, and $\rm C3$, which only guarantee the functionality of the LLM serving system, enforcing the SLO constraints~\eqref{eq:slo1}-\eqref{eq:slo3} requires additional hardware resources. This extra cost is the \textit{SLO tax}. Fig.~\ref{fig:insight2} illustrates the SLO tax by showing the minimum requirements derived from both the stability conditions and SLO constraints as $p_{\rm SLO}$ tightens. The evaluation adopts the GPT-3 175B model with A100 devices, $\lambda=20$ req/s, and $(\tau_{\rm pre},\tau_{\rm KV},\tau_{\rm dec})=(1.5,0.2,0.02)$ s. Both exponential light-tailed (${\rm CV} = 1$) and log-normal heavy-tailed (${\rm CV} \in \{1.5,2\}$) cases are considered.
	
	First, Fig.~\ref{fig:insight2}(a) shows the minimum $k_{\rm p}$ under the stability
	condition $\rho_{\rm p}<1$ and the SLO constraint~\eqref{eq:slo1} characterized by~\eqref{eq:qoslhs1}. It shows that $k_{\rm p}$ diverges rapidly as $p_{\rm SLO}\to 1$, and the heavy-tailed effect of input lengths significantly aggravates the hardware demand. 
	
	Besides, Fig.~\ref{fig:insight2}(b) shows the minimum bandwidth $B$ against $1-p_{\rm SLO}$, characterized by~\eqref{eq:kvlight} and~\eqref{eq:kvheavy} for light- and heavy-tailed workloads, respectively. According to~\eqref{eq:kvlight} and~\eqref{eq:kvheavy}, the SLO tax beyond the stability floor $B>\lambda g\gamma\ell_{\rm i}$ grows linearly with $\log\frac{1}{1-p_{\rm SLO}}$, with a slope proportional to $e^{\sigma_{\rm i}^2}/2=(1+c_{\rm TR}^2)/2$. In other words, the required bandwidth increases exponentially with $p_{\rm SLO}$, and the heavy-tailed effect further amplifies the exponential factor.
	
	Furthermore, Fig.~\ref{fig:insight2}(c) shows the minimum $k_{\rm d}$ against $p_{\rm SLO}$, where the SLO constraint follows from ${\rm C1}$ and the stability floor follows from ${\rm C3}$. As $p_{\rm SLO}$ increases, the quantile term $\theta_{\rm LT} P^{-1}_{\alpha_{\rm LT}}(p_{\rm SLO})$ or  $e^{\mu_\ell+\sigma_\ell\Phi^{-1}(p_{\rm SLO})}$ in~\eqref{eq:dec_ub1l} or~\eqref{eq:dec_ub1h} inflates the context length within a batch that must be served within TPOT threshold $\tau_{\rm dec}$, and the device count increases accordingly. As $p_{\rm SLO}\to 1$, the additional demand incurred by the SLO tax can reach about twice the stability condition, revealing the necessity of SLO-aware resource allocation.

	\subsection{Latency Budget Allocation}
	
	\begin{figure*}[t]
		\centering
		\includegraphics[width=0.8\textwidth]{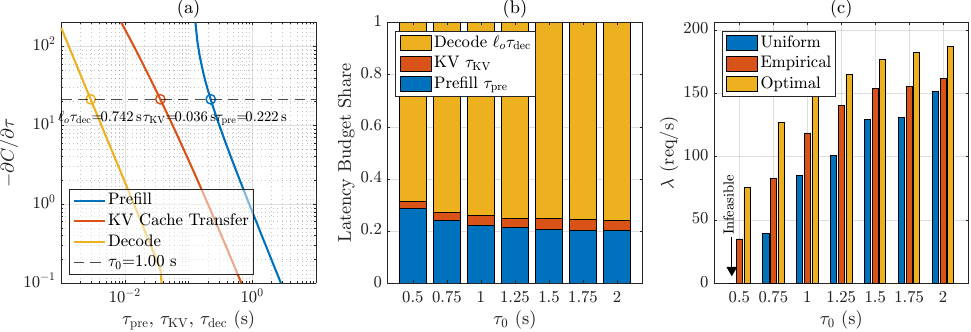}
		\caption{Latency budget $\tau_0$ allocation across the prefill, KV cache transfer, and decode stages. (a) Marginal benefits curves with varying $\boldsymbol{\tau}$, (b) optimal budget shares versus budget $\tau_0$, and (c) goodput versus budget $\tau_0$ compared with empirical and uniform baselines.}
		\label{fig:insight3}
	\end{figure*}
	
	We further investigate the latency thresholds in the SLO constraints~\eqref{eq:slo1}-\eqref{eq:slo3}. In agentic workflows, a single inference task may contain tens of user requests, and the completion time of every request directly affects the overall service latency. 
	Therefore, in addition to TTFT and TPOT, the aggregate latency $\tau_0$ is also an important metric for LLM services. For each request, the overall budget of the aggregate latency $\tau_0$ can be expressed as
	\begin{equation}
		\tau_0 \geq \tau_{\rm pre} + \tau_{\rm KV} + \ell_{\rm o} \tau_{\rm dec}.
		\label{eq:tau0decomp}
	\end{equation}
	To achieve goodput maximization, we need to allocate the latency budget $\tau_0$ to three stages. Since the deployment cost $C_{\rm max}$ increases monotonically with the achievable arrival rate $\lambda$, maximizing goodput is essentially equivalent to minimizing the deployment cost at a fixed $\lambda$. Hence, the allocation of $\tau_0$ is equivalent to minimizing the deployment cost
	\begin{equation}
		C(\boldsymbol{\tau}) = k_{\rm p}(\tau_{\rm pre}) C_{\rm p}+B(\tau_{\rm KV}) C_{\rm KV}+k_{\rm d}(\tau_{\rm dec})C_{\rm d},\label{eq:costfunc}
	\end{equation}
	subject to~\eqref{eq:tau0decomp}, where $\boldsymbol{\tau} = \{ \tau_{\rm pre},\tau_{\rm KV}, \tau_{\rm dec} \}$. The Lagrangian of the cost function~\eqref{eq:costfunc} is given by
	\begin{equation}
		\begin{aligned}
			\mathcal L(\boldsymbol{\tau})={}&k_{\rm p}(\tau_{\rm pre}) C_{\rm p}+B(\tau_{\rm KV}) C_{\rm KV}+k_{\rm d}(\tau_{\rm dec})C_{\rm d}\\
			&+\nu\left(\tau_{\rm pre}+\tau_{\rm KV}+\ell_{\rm o}\tau_{\rm dec}-\tau_0
			\right),
			\label{eq:tau0lag}
		\end{aligned}
	\end{equation}
	whose first-order optimal conditions yield
	\begin{equation}
		C_{\rm p} \frac{ \partial k_{\rm p} }{\partial\tau_{\rm pre}}
		=C_{\rm KV}\frac{\partial B}{\partial\tau_{\rm KV}}
		=\frac{C_{\rm d}}{\ell_{\rm o}}
		\frac{\partial k_{\rm d}}{\partial\tau_{\rm dec}}
		=-\nu.
		\label{eq:tau0kkt}
	\end{equation}
	Since the minimum cost of each stage is a non-increasing function of its own latency budget, the solution to~\eqref{eq:tau0kkt} is the unique global optimum. Therefore, the optimal latency budget allocation can be obtained by solving~\eqref{eq:tau0kkt} numerically. 
	The physical interpretation of~\eqref{eq:tau0kkt} is that the marginal benefit of latency budget allocation is identical across three stages. 
	
	Fig.~\ref{fig:insight3}(a) shows the marginal benefit curves $-\partial C/\partial\tau$ of three stages, where the optimal allocation at $\tau_0=1$ s is marked by the dashed line. It is validated that the marginal benefits $-\partial C/\partial\tau$ are decreasing over $\tau$, so the optimal budget allocation can be obtained via numerical methods. 
	
	Fig.~\ref{fig:insight3}(b) illustrates the latency budget allocation share among three stages. When the total budget $\tau_0$ is small, a larger fraction of the budget is allocated to the prefill stage. 
	This is because a tight TTFT threshold $\tau_{\rm pre}$ substantially increases the prefill queuing delay and the required number of prefill instances, resulting in a larger marginal cost reduction when $\tau_{\rm pre}$ is relaxed. As $\tau_0$ increases, the prefill resource requirement gradually approaches its stability floor, while the KV cache transfer and decode requirements remain sensitive to their latency thresholds, so the latency shares these stages keep increasing.
	
	Fig.~\ref{fig:insight3}(c) shows the goodput achieved by the optimal allocation, where an empirical latency budget allocation with $25\%:25\%:50\%$ share split and a uniform share split baselines are considered.
	The optimal allocation consistently attains the highest
	goodput, and the gain is most pronounced under tight budgets. 
	As $\tau_0$ increases, the goodput of all schemes increases, while the proposed allocation scheme still achieves the highest goodput performance.

	\begin{table}[t]
		\centering
		\caption{Basic Parameters of the Evaluated LLMs.}
		\label{tab:models}
		\renewcommand{\arraystretch}{1.15}
		\footnotesize
		\setlength{\tabcolsep}{4pt}
		\begin{tabular}{@{}lccc@{}}
			\hline\hline
			\textbf{Parameter} & \textbf{LLaMA 3.1 8B} & \textbf{Qwen2.5 32B} & \textbf{GPT-3 175B} \\
			\hline
			Attention Mechanism & GQA & GQA & MHA \\
			$N_{\rm mod}$ & $4{,}096$ & $5{,}120$ & $12{,}288$ \\
			$N_{\rm head}$ & $32$ & $40$ & $96$ \\
			$N_{\rm KV}$ & $8$ & $8$ & $96$ \\
			$N_{\rm attn}$ & $128$ & $128$ & $128$ \\
			$N_{\rm hid}$ & $14{,}336$ & $27{,}648$ & $49{,}152$ \\
			$N_{\rm layer}$ & $32$ & $64$ & $96$ \\
			$g=N_{\rm KV}/N_{\rm head}$ & $0.25$ & $0.2$ & $1$ \\
			$\gamma$ (GiB/token) & $4.883\!\times\!10^{-4}$ & $1.221\!\times\!10^{-3}$ & $4.395\!\times\!10^{-3}$ \\
			$\Omega_{\rm LLM}$ (GiB) & $14.90$ & $59.60$ & $325.96$ \\
			\hline\hline
		\end{tabular}
	\end{table}
	
	\section{Main Results}
	
	In this section, we validate the modeling accuracy of the proposed closed-form approximations and evaluate the performance of the proposed resource allocation scheme. 
	\subsection{Parameter Setup}

	Unless otherwise stated, in the following evaluation part, we conduct LLM inference experiments using the high-performance SGLang framework~\cite{3737916.3739916} on multiple NVIDIA A100 GPU devices. 
	In the simulation, the cost for each device is set to $C_{\rm p} = C_{\rm d} = 5$ USD per hour, and the cost for bandwidth is set to $C_{\rm KV} = 0.1$ USD per GiB per hour. The SLO constraint probability is $p_{\rm SLO} = 0.95$, and the typical latency thresholds $\tau_{\rm pre}$, $\tau_{\rm KV}$, and $\tau_{\rm dec}$ are set to $0.5$, $0.15$, and $0.04$ seconds, respectively. 
	$p_{\rm mem}$ and $p_{\rm join}$ are both set to $0.99$. 
	We mainly consider user requests with heavy-tailed workload, where $L_{\rm i}$ follows a log-normal distribution as $L_{\rm i} \sim \mathcal{LN}[\mu_{\rm i},\sigma_{\rm i}^2]$ with mean $\ell_{\rm i} = \mathbb{E}[L_{\rm i}]=e^{\mu_{\rm i} + \sigma_{\rm i}^2/2} = 1,024$ and CV $\sigma_{\rm i}/\mu_{\rm i} = 1.25$~\cite{316730}. 
	The output length follows an exponential distribution $L_{\rm o} \sim {\rm Exp} [1/\ell_{\rm o}]$ with $\ell_{\rm o} = 256$. 
	
	\begin{figure}[t]
		\centering
		\includegraphics[width=0.45\textwidth]{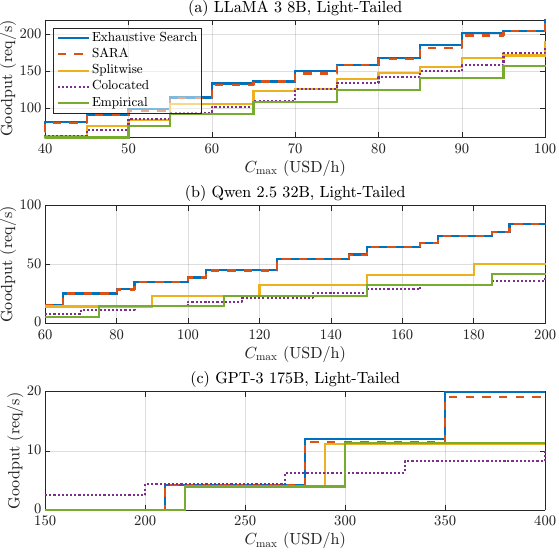}
		\caption{Goodput performance under light-tailed workload and varying cost budget $C_{\rm max}$ for (a) LLaMA 3 8B, (b) Qwen2.5 32B, and (c) GPT-3 175B models.}
		\label{fig:glight}
	\end{figure}
	
	The baseline models cover a wide range of memory footprints, including LLaMA 3.1 8B~\cite{grattafiori2024llama3herdmodels}, Qwen2.5 32B~\cite{qwen2025qwen25technicalreport}, and GPT‑3 175B~\cite{NEURIPS2020_1457c0d6}. The parameters of these baseline models are illustrated in Table~\ref{tab:models}.

	\subsection{Resource Allocation}

	\begin{figure}[t]
		\centering
		\includegraphics[width=0.45\textwidth]{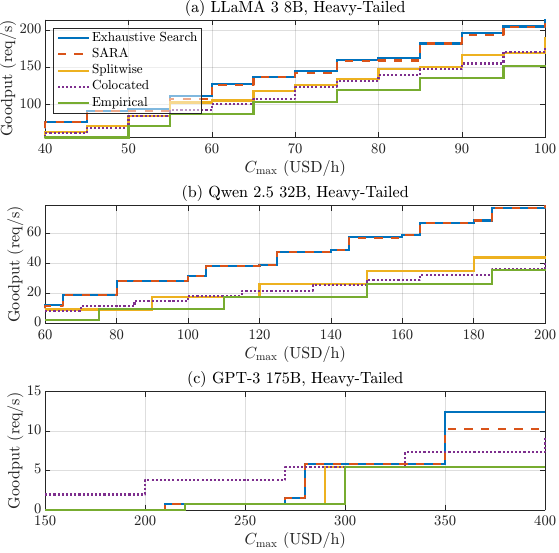}
		\caption{Goodput performance under heavy-tailed workload and varying cost budget $C_{\rm max}$ for (a) LLaMA 3 8B, (b) Qwen2.5 32B, and (c) GPT-3 175B models.}
		\label{fig:gheavy}
	\end{figure}
	
	We first evaluate the effectiveness of the proposed SLO-aware resource allocation, i.e., SARA, scheme in maximizing goodput under a fixed deployment cost. The baseline methods are as follows.
	\begin{itemize}
		\item \textbf{Exhaustive Search}: Enumerates all feasible configurations of $\{ k_{\rm p}, B,k_{\rm d}, N_{\rm bat} \}$ and selects the configuration that provides the best goodput performance under the given cost $C_{\rm max}$ and SLO constraints~\eqref{eq:slo1}-\eqref{eq:slo3}. %
		\item \textbf{Empirical}: The overall cost $C_{\rm max}$ is allocated with fixed proportions among the three stages. In this work, the proportion is set as $45\%:10\%:45\%$. 
		\item \textbf{Splitwise}: Apart from the prefill and decode resource pools, this scheme considers a mixed resource pool where the short user requests are routed to. Prefill and decode stages are operated together within the mixed resource pool. 
		\item \textbf{Colocated}: Only one mixed pool is considered in this scheme, where the prefill and decode stages operate concurrently. The latency introduced by KV cache transfer is eliminated, while the interference between the two stages persists.
	\end{itemize}
	
	Figs.~\ref{fig:glight} and~\ref{fig:gheavy} show the achievable goodput under light-tailed and heavy-tailed workloads, respectively. The goodput increases with the deployment cost for all methods because additional deployment costs lead to higher hardware capacity to support a higher request rate. For most of the cases, the proposed SARA method achieves the highest goodput performance, preceded only by the exhaustive search scheme, which reveals the superiority of the SARA scheme. For extremely large models, e.g., GPT-3 175B, with a limited cost budget $C_{\rm max}$, the colocated scheme achieves slightly higher goodput performance while all the disaggregated schemes fail to operate. 
	Note that the goodput improvement becomes more significant for heavy-tailed workloads and higher cost budgets, showing the effectiveness of the proposed scheme in agentic workflows and large-scale deployments.
	
	\begin{figure}[t]
		\centering
		\includegraphics[width=0.45\textwidth]{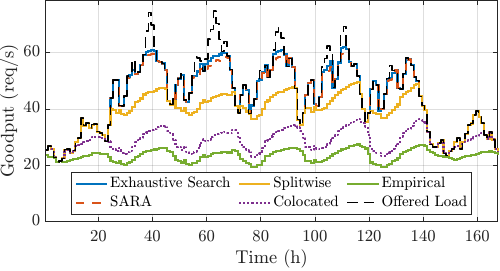}
		\caption{Goodput performance of the replayed seven-day trace~\cite{10946802} with the LLaMA 3.1 8B model on A100 GPUs.}
		\label{fig:replay}
	\end{figure}
	
	{We then replay the seven-day public Microsoft Azure LLM inference trace~\cite{10946802} to evaluate the allocation schemes under a time-varying production workload. We employ the LLaMA 3.1 8B model with overall cost $C_{\rm max} = 35$ USD/h on A100 GPUs. The resource allocation algorithms are executed once per hour, and the goodput results are shown in Fig.~\ref{fig:replay}. 
	During off-peak hours, the goodput difference across the algorithms is marginal. However, pronounced differences can be observed during peak hours. The exhaustive search achieves the highest goodput performance and is tightly tracked by the proposed SARA scheme. The maximum performance gap between the two schemes is $3.8\%$ under extreme workloads. On the contrary, the splitwise, colocated, and empirical schemes fail to utilize resources efficiently, and the maximum performance degradations reach $21.5\%$, $45.6\%$, and $57.2\%$ during peak periods, respectively. These results demonstrate that the proposed scheme achieves superior resource utilization over the baselines.
	}

	\subsection{Computational Complexity}
	\begin{table*}[t]
		\centering
		\caption{Execution Time of Resource Allocation Schemes (seconds).}
		\label{tab:exetime}
		\renewcommand{\arraystretch}{1.15}
		\footnotesize
		\setlength{\tabcolsep}{4pt}
		\begin{tabular}{@{}l c c c c c c@{}}
			\hline\hline
			& \multicolumn{3}{c}{\textbf{Light-Tailed}} 
			& \multicolumn{3}{c}{\textbf{Heavy-Tailed}} \\
			\cline{2-4}\cline{5-7}
			\textbf{Method}
			& \textbf{LLaMA 3.1 8B}
			& \textbf{Qwen2.5 32B}
			& \textbf{GPT-3 175B}
			& \textbf{LLaMA 3.1 8B}
			& \textbf{Qwen2.5 32B}
			& \textbf{GPT-3 175B} \\
			\hline
			Exhaustive Search
			& $1.44\times 10^3$ & $2.44\times 10^3$ & $6.19\times 10^2$
			& $1.44\times 10^3$ & $2.44\times 10^3$ & $6.19\times 10^2$ \\
			SARA
			& $\bf 0.70$ & $\bf 1.60$ & $\bf 0.61$
			& $\bf 0.44$ & $\bf 1.11$ & $\bf 1.02$ \\
			Splitwise
			& $60.9$ & $78.6$ & $2.02$
			& $60.8$ & $78.5$ & $2.01$ \\
			Colocated
			& $13.0$ & $7.71$ & $2.46$
			& $13.0$ & $7.73$ & $2.47$ \\
			\hline\hline
		\end{tabular}
	\end{table*}
	
	We also investigate the computational complexity of the proposed resource allocation algorithm, and the results are shown in Table~\ref{tab:exetime}. Exhaustive search requires extremely high execution time as it enumerates millions of hardware combinations to find the optimal configuration. Both Splitwise and colocated schemes estimate the required resources and prune the search space, so the allocation search overheads are substantially reduced. 
	Empirical allocation follows a resource allocation preset, and therefore exhibits no execution time at the cost of lower goodput performance. 
	The proposed SARA completes the allocation with no more than $1.60$ s, achieving an acceleration of several orders of magnitude over exhaustive search, and is faster than all baseline methods, especially when the resource requirement is sufficiently large. 
	Combined with the goodput results in Figs.~\ref{fig:glight} and~\ref{fig:gheavy}, these results demonstrate that SARA is suitable for practical resource planning and system reconfiguration. 
	{It is worth noting that the execution time in Table~\ref{tab:exetime} drops significantly for the GPT-3 175B model. This is because its large memory footprint constrains the feasible solution space under limited resources. Meanwhile, the proposed SARA scheme consistently executes in around $1$~s, as it reduces the solution space via closed-form expressions, eliminating the need for exhaustive enumeration.}

	\section{Conclusion}
	
	In this paper, we developed an SLO-aware resource allocation framework for disaggregated LLM serving systems. We modeled the prefill, KV cache transfer, and decode stages using an M/G/$k_{\rm p}$ queue, an M/G/$1$ queue, and a generalized birth-death process, respectively. Based on these models, we derived tractable latency quantile characterizations and resource requirements under both light-tailed and heavy-tailed workloads. 
	{Based on the proposed analytical framework, we found and validated several insights on data center planning and resource allocation. First, the saturated token throughput of the decode stage depends on the HBM bandwidth and the LLM parameters, and enlarging the batch size without violating the TPOT SLO helps improve the overall throughput. Besides, the required numbers of prefill and decode devices scale quadratically with the input and output lengths, revealing the fundamental resource pressure imposed by long-context agentic workloads. Furthermore, both the SLO constraints and the heavy-tailed workload significantly amplify the resource requirements, imposing hardware demands well beyond stability conditions. Finally, both the proposed latency budget allocation and the SARA schemes effectively improve the system goodput with low complexity, demonstrating the practicality of the proposed analytical framework.}

	\bibliographystyle{IEEEtran}
	\bibliography{IEEEabrv,references}
	\clearpage
	\appendices
	\section{Latency Dominant Factors}
	\label{sec:domi}
	According to~\eqref{eq:prefillTp} and~\eqref{eq:decodetime}, the service time in the prefill and decode stage is determined by multiple system parameters. To facilitate efficient analysis of $\mathcal{P}_1$, in this subsection, we first identify the dominant performance parameters under practical configurations.
	\subsection{Compute-Bound Prefill Service Time}
	In the prefill stage, according to~\eqref{eq:Tp1} and~\eqref{eq:Tp2}, the prefill service time $T_{\rm p}$ is determined by the compute capacity $F_{\rm p}$ and HBM bandwidth $B_{\rm HBM}$. %
	To further investigate the dominant factors of service time $T_{\rm p}$, we first solve $T_{\rm p,1}(L_{\rm i})>T_{\rm p,2}(L_{\rm i})$, which yields a quadratic form of $L_{\rm i}$ as
	\begin{equation}
		aL_{\rm i}^2 + b L_{\rm i} +c > 0
		\label{eq:prefillthres}
	\end{equation}
	with $b = B_{\rm HBM} N_{\rm layer} N_{\rm mod}\left( 1+2N_{\rm hid}+2(1+g)N_{\rm mod} \right) - F_{\rm p}g\gamma$, $a = B_{\rm HBM} N_{\rm layer} N_{\rm mod}$, and $c = -F_{\rm p} \Omega_{\rm LLM}$. For practical LLM serving systems\footref{fn:1}, 
	the quadratic term $aL_{\rm i}^2$ is negligible compared with the linear term $bL_{\rm i}$ as
	\begin{align}
		\frac{aL_{\rm i}^{2}}{bL_{\rm i}}
		&=\frac{L_{\rm i}}{B_{\rm HBM} N_{\rm layer} N_{\rm mod}\left( 1\!+\!2N_{\rm hid}\!+\!2(1\!+\!g)N_{\rm mod} \right) \!-\! F_{\rm p}g\gamma}\notag\\
		&\approx 0.004
		\ll 1.
	\end{align}
	Hence,~\eqref{eq:prefillthres} can be approximately reduced to a linear inequality $bL_{\rm i}+c > 0$, which yields the length threshold
	\begin{equation}
		\tilde{L}_{\rm i} %
		= \frac{F_{\rm p} \Omega_{\rm LLM} }{B_{\rm HBM} N_{\rm layer} N_{\rm mod}\left( 1\!+\!2N_{\rm hid}\!+\!2(1\!+\!g)N_{\rm mod} \right) \!-\! F_{\rm p}g\gamma },\notag%
	\end{equation}
	above which~\eqref{eq:prefillthres} holds. 
		Based on practical parameter setups\footref{fn:1}, the length threshold is around $\tilde{L}_{\rm i} \approx 156$ tokens, as shown in Fig.~\ref{fig:li}(a). Due to prompt prefixes and accumulated input and output context in multi-turn conversations of LLM user requests, the input token length $L_{\rm i}$ 
	in realistic scenarios is typically {around thousands of tokens, which is}
	much longer than $\tilde{L}_{\rm i}$~\cite{patel2024splitwise,10946802,316730}. Consequently, the prefill service time is mainly determined by compute capacity, i.e., $T_{\rm p} \approx T_{\rm p,1}$. 
	
	\subsection{HBM Bandwidth-Bound Decode Service Time}
\begin{figure}[t]
		\centering
		\includegraphics[width=0.45\textwidth]{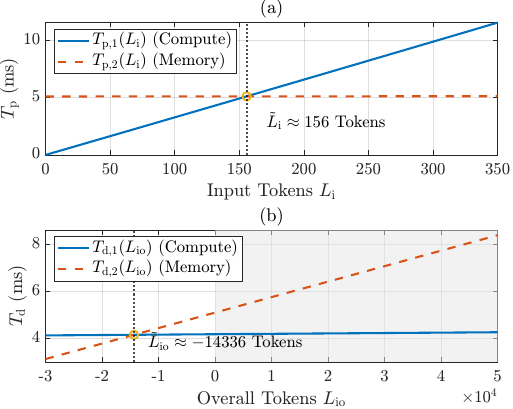}
		\caption{Token thresholds for (a) prefill and (b) decode stages under practical parameters stated in Footnote~\ref{fn:1}.}
		\label{fig:li}
	\end{figure}

	Similarly, we first calculate $T_{\rm d,1}(L_{\rm io})<T_{\rm d,2}(L_{\rm io})$ according to~\eqref{eq:Td1} and~\eqref{eq:Td2}, which yields
	\begin{align}
		&\left( g\gamma F_{\rm d} -  2B_{\rm HBM}N_{\rm layer} N_{\rm mod}\right)L_{\rm io}\\
		>{}&2B_{\rm HBM}N_{\rm layer}N_{\rm bat}\!\left( (1\!+\!g)N_{\rm mod}^2\!+\!N_{\rm mod}N_{\rm hid} \right)\!-\!F_{\rm d}\Omega_{\rm LLM}.\notag
	\end{align}
	Since $g\gamma F_{\rm d} -  2B_{\rm HBM}N_{\rm layer} N_{\rm mod}>0$ typically holds true under practical parameters\footref{fn:1}, 
	the threshold of $L_{\rm io}$ is given by
	\begin{align}
		&\tilde{L}_{\rm io}\\
		=&\frac{2B_{\rm HBM}N_{\rm layer}N_{\rm bat}\!\left( (1\!+\!g)N_{\rm mod}^2\!+\!N_{\rm mod}N_{\rm hid} \right)\!-\!F_{\rm d}\Omega_{\rm LLM}}{F_{\rm d}g\gamma - 2B_{\rm HBM}N_{\rm layer} N_{\rm mod}}.\notag
	\end{align}
	The resource bottleneck is HBM bandwidth $B_{\rm HBM}$ for $L_{\rm io}\ge \tilde{L}_{\rm io}$, and is compute capacity $F_{\rm d}$ otherwise. 
	With practical batch size such as $N_{\rm bat} = 128$, the token threshold  $\tilde{L}_{\rm io}$ is even negative as shown in Fig.~\ref{fig:li}(b),  
	so the decode service time is mainly determined by the HBM bandwidth $B_{\rm HBM}$, i.e., $T_{\rm d} \approx T_{\rm d,2}$.
	
	\section{Heavy-Tailed Analysis for KV Cache Transfer}
	\label{append:2}
	According to Kingman-type heavy traffic approximation~\cite{kingman1992poisson,whitt2002stochastic}, we resort to analyze the waiting time as
	\begin{equation}
		\Pr\left(T_{\rm TR}^{\rm W}\geq t\right)\approx \rho_{\rm KV} e^{-\frac{t}{w_{\rm TR}}},
		\label{eq:kv_ht_waiting}
	\end{equation}
	where 
	\begin{equation}
		w_{\rm TR}
		= \frac{1+c_{\rm TR}^2}{2} \frac{\rho_{\rm KV}}{\lambda\left( 1-\rho_{\rm KV}\right)}
		=
		\frac{e^{\sigma_{\rm i}^2}}{2} \frac{g\gamma \ell_{\rm i} }{B-\lambda g\gamma \ell_{\rm i} }\notag
		\label{eq:kv_ht_scale}
	\end{equation}
	is the modified tail decay factor, $\ell_{\rm i} = \mathbb{E}[L_{\rm i}] = e^{\mu_{\rm i}+{\sigma_{\rm i}^2}/{2}}$, and $c_{\rm TR} = \sqrt{e^{\sigma_{\rm i}^2}-1}$ is the CV of $T_{\rm TR}$. Under high utilization $\rho_{\rm KV}\to 1$, the waiting time $T_{\rm TR}^{\rm W}$ asymptotically dominates the sojourn time as~\cite{Pollaczek1930,Abate1994}
	\begin{equation}
		\lim_{\rho_{\rm KV}\to 1}\frac{\mathbb E[T_{\rm TR}^{\rm W}]}
		{\mathbb E[T_{\rm TR}]} = \lim_{\rho_{\rm KV}\to 1} \frac{\rho_{\rm KV}\left( 1+c_{\rm TR}^2 \right)}{2(1-\rho_{\rm KV})}\gg 1,\label{eq:meankv}
	\end{equation}
	and
	\begin{equation}
		\!\!\lim_{\rho_{\rm KV}\to 1}\frac{ {\rm Var}(T_{\rm TR}^{\rm W})}
		{ {\rm Var}(T_{\rm TR})} = \!\! \lim_{\rho_{\rm KV}\to 1} \frac{\frac{\rho_{\rm KV} e^{3\sigma_{\rm i}^2}}{3(1-\rho_{\rm KV})}+\frac{\rho_{\rm KV}^2 e^{2\sigma_{\rm i}^2}}{4(1-\rho_{\rm KV})^2}}{e^{\sigma_{\rm i}^2}-1}\gg 1.\label{eq:varkv}
	\end{equation}
	Eqs.~\eqref{eq:meankv} and~\eqref{eq:varkv} indicate that $T_{\rm TR}$ is a random variable with much smaller expectation value and variance than $T_{\rm TR}^{\rm W}$. 
	Hence, we can safely approximate the tail behavior of sojourn latency using the waiting time $T_{\rm TR}^{\rm W}$ as 
	\begin{equation}
		T_{\rm KV} \approx T_{\rm TR}^{\rm W},%
	\end{equation}
	which simplifies the analytical framework without incurring noticeable error. Hence, the SLO constraint~\eqref{eq:slo2} is characterized as
	\begin{equation}
		\tau_{\rm KV}\geq w_{\rm TR}\log\frac{\rho_{\rm KV}}{1-p_{\rm SLO}},%
	\end{equation}
	which yields
	\begin{equation}
		B\geq \lambda g\gamma \ell_{\rm i} + \frac{e^{\sigma_{\rm i}^2}}{2}\frac{g\gamma \ell_{\rm i}}{\tau_{\rm KV}}\log\frac{\rho_{\rm KV}}{1-p_{\rm SLO}}.%
	\end{equation}
	\section{Derivation of Condition $\rm C3$}
	\label{append:3}
	Statistically, the output length $L_{\rm o}^{(n)}$ follows an exponential distribution as $L_{\rm o}^{(n)}\sim {\rm Exp}[1/\ell_{\rm o}]$, and is independent with the input length $L_{\rm i}^{(n)}$~\cite{10946802,316730,ModServe}. Therefore, in each decoding iteration, the probability of reaching the EOS token for each request is
	\begin{equation}
		p_0 = \Pr\left( L_{\rm o}\leq n+1\mid L_{\rm o}\geq n \right) = 1-e^{-\frac{1}{\ell_{\rm o}}}.
	\end{equation}
	The number of departing requests $D_k$ then follows a binomial distribution $D_k\sim {\mathcal B}[ N_{\rm bat}^{(k)}, p_0]$. Let the number of arrivals in the $k$-th decoding iteration be $A_k \sim {\rm Poisson}[\lambda  T_{\rm d}]$, so the number of requests to be processed in the $(k+1)$-th iteration is
	\begin{equation}
		N_{\rm bat}^{(k+1)} = N_{\rm bat}^{(k)} - D_k + A_k.\label{eq:overall_proc0}
	\end{equation}
	At the steady state of $N_{\rm bat}^{(k)}$, i.e., $k\to \infty$, we should have 
	\begin{equation}
		\lim_{k\to\infty} \left( \mathbb{E}\left[N_{\rm bat}^{(k+1)} \right] -\mathbb{E}\left[ N_{\rm bat}^{(k)} \right] \right) = 0\label{eq:steadystate}
	\end{equation}
	to ensure the stability of the decoding system. 
	In this case, taking the conditional expectation on both sides of~\eqref{eq:overall_proc0}, we have
	\begin{equation}
		\!\!\!\!
		\begin{aligned}
			&\mathbb{E}\left[ N_{\rm bat}^{(k+1)}\mid N_{\rm bat}^{(k)} \right] \\={}& \mathbb{E}\left[ N_{\rm bat}^{(k)} - D_{k}\mid N_{\rm bat}^{(k)} \right] + \mathbb{E}\left[ A_{k}\mid N_{\rm bat}^{(k)} \right]\\
			={}& N_{\rm bat}^{(k)} e^{-\frac{1}{\ell_{\rm o}}} + \frac{\lambda \left( \Omega_{\rm LLM} + g\gamma N_{\rm bat}^{(k)}\left( \ell_{\rm i}+\ell_{\rm o} \right) \right)  }{k_{\rm d} B_{\rm HBM}}, %
		\end{aligned}
		\label{eq:condmean}
	\end{equation}
	where $\ell_{\rm i} = \mathbb{E}[L_{\rm i}]$. Taking the conditional mean over $N_{\rm bat}^{(k)}$ on both sides of~\eqref{eq:condmean} yields
	\begin{equation}
		\mathbb{E}\!\left[ N_{\rm bat}^{(k+1)} \right]\!=\! \mathbb{E}\!\left[ N_{\rm bat}^{(k)} \right]\!e^{-\frac{1}{\ell_{\rm o}}} \!+\! \frac{\lambda\! \left(\! \Omega_{\rm LLM} \!+\! g\gamma \mathbb{E}\!\left[ N_{\rm bat}^{(k)} \right]\!\left( \ell_{\rm i}\!+\!\ell_{\rm o} \right)\! \right)\!  }{k_{\rm d} B_{\rm HBM}}.
		\label{eq:uncondmean}
	\end{equation}
	Substituting~\eqref{eq:steadystate} into~\eqref{eq:uncondmean}, we obtain the steady state mean
	\begin{equation}
		\lim_{k\to \infty}\mathbb{E}\left[N_{\rm bat}^{(k)}\right] = \frac{\lambda \Omega_{\rm LLM}}{k_{\rm d}B_{\rm HBM} p_0 - \lambda g\gamma\left( \ell_{\rm i} +\ell_{\rm o} \right)}%
	\end{equation}
	subject to $k_{\rm d}B_{\rm HBM} p_0 > \lambda g\gamma\left(\ell_{\rm i} + \ell_{\rm o}\right)$, which further yields
	\begin{equation}
		\!\!{\rm C3}:~N_{\rm bat} >  \frac{\lambda \Omega_{\rm LLM}}{k_{\rm d}B_{\rm HBM} p_0 - \lambda g\gamma\left( \ell_{\rm i} +\ell_{\rm o} \right)}.%
	\end{equation}
	\section{Derivation of Condition $\rm C4$}
	\label{append:4}
	Following the notation in Appendix~\ref{append:3}, the conditional variance of $N_{\rm bat}^{(k+1)}$ is given by
	\begin{equation}
		\!\!\!\!
		\begin{aligned}
			&{\rm Var}\left( N_{\rm bat}^{(k+1)}\mid N_{\rm bat}^{(k)} \right)\\ 
			={}& {\rm Var}\left( N_{\rm bat}^{(k)}-D_k\mid N_{\rm bat}^{(k)} \right) + {\rm Var}\left( A_k\mid N_{\rm bat}^{(k)} \right)\\
			={}& N_{\rm bat}^{(k)} p_0 (1-p_0) + \frac{\lambda \left(  \Omega_{\rm LLM} +  g\gamma N_{\rm bat}^{(k)}\left( \ell_{\rm i}+\ell_{\rm o} \right) \right)}{k_{\rm d}B_{\rm HBM}}.%
		\end{aligned}
	\end{equation} 
	According to the law of total variance, we have
	\begin{equation}
		\begin{aligned}
			{\rm Var}\left(N_{\rm bat}^{(k+1)}\right) ={}& \mathbb{E}\left[{\rm Var}\left(N_{\rm bat}^{(k+1)}\mid N_{\rm bat}^{(k)}\right)\right]\\ 
			&+ {\rm Var}\left(\mathbb{E}\left[N_{\rm bat}^{(k+1)}\mid N_{\rm bat}^{(k)}\right]\right),
		\end{aligned}
	\end{equation}
	and the variance of steady state $N_{\rm bat}^{(k)}$ is given by%
	\begin{equation}
		\begin{aligned}
			&{\rm Var}\left(N_{\rm bat}^{(\infty)}\right) =\\& \frac{\displaystyle \frac{\lambda \Omega_{\rm LLM} }{ k_{\rm d}B_{\rm HBM}} +\left( p_0(1-p_0) + \frac{\lambda g\gamma ( \ell_{\rm i}+\ell_{\rm o} ) }{ k_{\rm d}B_{\rm HBM}} \right)\mathbb{E}\left[N_{\rm bat}^{(\infty)}\right] }{\displaystyle 1-\left( 1-p_0 +\frac{\lambda g\gamma ( \ell_{\rm i}+\ell_{\rm o} ) }{ k_{\rm d}B_{\rm HBM}} \right)^2}
			\label{eq:var1}
		\end{aligned}
	\end{equation}

	Note that given a practically large number of parallel requests within the batch, the steady state distribution of $N_{\rm bat}$ is the sum of i.i.d. Bernoulli $D_k$ and Poisson $A_k$ components. Hence, if there is no limit on the maximum batch size, the distribution of requests within the batch is approximately Gaussian according to the central limit theorem (CLT), i.e., $N_{\rm bat} \sim \mathcal{N}(\mu_{\rm bat}^{}, \sigma_{\rm bat}^2)$, where $\mu_{\rm bat} = \mathbb{E}[N_{\rm bat}^{(\infty)}]$ and $\sigma_{\rm bat}^2 = {\rm Var}(N_{\rm bat}^{(\infty)})$. 
	Therefore, we can empirically assign a large batch size to avoid excessive waiting time in the decode stage by
	\begin{equation} 
		{\rm C4}:~~N_{\rm bat}\geq \mu_{\rm bat} + \sigma_{\rm bat}\Phi^{-1}(p_{\rm join}),
	\end{equation}
	where $p_{\rm join}$ is the probability of directly joining the decoding batch without waiting. 
\end{document}